\documentclass[apj,iop,revtex4]{emulateapj}
\usepackage{natbib}
\usepackage{hyperref}
\usepackage{color}
\usepackage{xcolor}
\usepackage{amsmath}
\newcommand{\mytilde}{\raise.19ex\hbox{$\scriptstyle\sim$}}

\newcommand{\Rmnum}[1]{\expandafter\@slowromancap\romannumeral #1@}
\newcommand{\bvec}[1]{\textbf{\textit{#1}}}

\shorttitle{Weak-lensing Study of A115}
\shortauthors{Kim et al. (2018)}

\begin{document}

\title{Multi-wavelength Analysis of the Merging Galaxy Cluster A115}

\author{Mincheol Kim\altaffilmark{1}, M. James Jee\altaffilmark{1,2}, Kyle Finner\altaffilmark{1}, \\ Nathan Golovich\altaffilmark{3}, David M. Wittman\altaffilmark{2},
R. J. van Weeren\altaffilmark{4}, W. A. Dawson\altaffilmark{3}}
\altaffiltext{1}{Yonsei University, Department of Astronomy, Seoul, Korea; chul542@yonsei.ac.kr, mkjee@yonsei.ac.kr}
\altaffiltext{2}{Department of Physics, University of California, Davis, California, USA}
\altaffiltext{3}{Lawrence Livermore National Laboratory, 7000 East Avenue,Livermore, CA 94550, USA}
\altaffiltext{4}{Leiden Observatory, Leiden University, P.O. Box 9513, 2300
RA Leiden, the Netherlands}

\begin{abstract}
A115 is a merging galaxy cluster at $z\sim0.2$ with a number of remarkable features including a giant ($\mytilde2.5$~Mpc) radio relic, two asymmetric X-ray peaks with trailing tails, and a peculiar line-of-sight velocity structure. We present a multi-wavelength study of A115 using optical imaging data from Subaru, X-ray data from {\it Chandra}, and spectroscopic data from the Keck/DEIMOS and MMT/Hectospec instruments. Our weak-lensing analysis shows that the cluster is comprised of two subclusters whose mass centroids are in excellent agreement with the two BCG positions ($\lesssim10\arcsec$). By modeling A115 with a superposition of two Navarro-Frenk-White halos, we determine the masses of the northern and southern subclusters to be $M_{200}=1.58_{-0.49}^{+0.56}\times 10^{14} M_{\sun}$ and $3.15_{-0.71}^{+0.79}\times 10^{14} M_{\sun}$, respectively. Combining the two halos, we estimate the total cluster mass to be $M_{200}=6.41_{-1.04}^{+1.08}\times10^{14} M_{\sun}$ at $R_{200}=1.67_{-0.09}^{+0.10}$~Mpc. These weak-lensing masses are significantly (a factor of 3--10) lower than what is implied by the X-ray and optical spectroscopic data. We attribute the difference to the gravitational and hydrodynamic disruption caused by the collision between the two subclusters.
\end{abstract}


\section{Introduction}
\label{sec_introduction}
Merging galaxy clusters are rich in astrophysical processes. Gravitational interaction distorts the dynamical structure of the pre-merger halos. Coulomb interaction leads, for example, to ram pressure stripping, plasma heating, and shock propagation. If dark matter particles interact non-gravitationally, the merger may produce measurable offsets between galaxies and weak-lensing mass peaks \citep{Markevitch2004,Randall2008}. Therefore, studying merging galaxy clusters in detail with observations and numerical simulations enables us to refine our knowledge on these astrophysical processes and possibly probe fundamental physics.

However, interpretation of observations of merging clusters is difficult. They provide only a single snapshot in the long merger history, which does not provide sufficient information to 
differentiate merging scenarios. Multi-wavelength observations aide in resolving the degeneracy. For example, a presence of radio relics is a strong indication that the intracluster medium (ICM) has already experienced significant Coulomb interactions and developed shocks \citep{Ferrari2008,bruggen2011,Vazza2012,Skillman2013}. The orientation and location of the relics
provide constraints on the merger axis. In addition, measurements of the spectral index and its steepening enable us to obtain Mach numbers of the shock, which is crucial for inferring the collision velocity \citep[e.g.][]{bonafede2014, Stroe2014,Urdampilleta2018,Gennaro2018,Hoang2018}.
The morphology of the X-ray emission and its offset with respect to galaxies can help us to estimate the direction of motion of the substructure because ICM is subject to ram pressure while galaxies are effectively collisionless.
X-ray temperature maps provide invaluable information on the dynamical state of the ICM such as shock-induced heating. Optical and near-IR spectroscopic data reveal exclusive information on the line-of-sight (LOS) velocity structure of the system and aide in our estimation of the merger geometry when combined with other velocity constraints \citep[e.g.,][]{Monteiro-Oliveira2017}. Finally, weak-lensing studies inform us of the dark matter distribution of the merging system and allow us to quantify the mass of each merging component \citep[e.g.,][]{Ragozzine2012,Soucail2012,Jee2015,Jee2016,Finner2017}.
%Optical and near-IR spectroscopic data reveal exclusive information on the line-of-sight (LOS) velocity structure of the system and are useful to determine the merger axis angle with respect to the plane of the sky \citep[e.g.,][]{Golovich2017}
Despite the consensus that merging galaxy clusters are useful astrophysical laboratories, the numerical simulation of radio relics is in its infancy. The major difficulty is our lack of understanding on how merger shocks lead to such powerful acceleration of electrons to relativistic speeds enabling luminous synchrotron emission. Because shocks alone cannot achieve such high efficiency, currently the so-called re-acceleration model is receiving a growing attention \citep[e.g.,][]{Kang2011,Kang2012,Pinzke2013,Kang2015}. That is, existing fossil electrons seeded by nearby active galactic nuclei or radio galaxies are
re-accelerated to relativistic speeds by ICM shocks triggered by cluster mergers. To date, there are only a few merging systems that show direct evidence for this re-acceleration scenario \citep[e.g.][]{bonafede2014, vanWeeren2017}.

In this paper, we present a multi-wavelength study of Abell~115 (hereafter A115), one of the few systems that have been considered as a test case to constrain the origin of the shock-relic connection with the re-acceleration model. In general, it is believed that a radio relic becomes observable when a merger happens nearly in the plane of the sky under the hypothesis that the merger shock propagates as a form of shallow spherical shell along the merger axis \citep[e.g.,][]{Golovich2017}. A115 is an X-ray luminous cluster with a distinct binary morphology \citep{Forman1981}. The northern X-ray peak (hereafter A115N) hosts a cool core and is much brighter in X-ray emission than the southern peak (hereafter A115S). The asymmetric X-ray morphology and its trailing feature indicate that A115N is moving southwest and the gas is being stripped.
A115S, separated by $\mytilde900$ kpc from A115N, is
hotter but less bright in X-ray. Similarly to A115N, the disturbed X-ray morphology of A115S has been attributed to its motion to the northeast. Thus, one quick interpretation of the X-ray observation and the presence of the radio relic is that A115 is a post-merger binary cluster with the two subclusters orbiting around each other nearly on the plane of the sky. However, many lines of evidence suggest that A115 is a much more complex system than this simplistic picture. Based on their 88 spectroscopic members, \cite{Barrena2007} claim that the line of sight (LOS) velocity difference between A115N and A115S is very large ($\mytilde1600~\mbox{km}~\mbox{s}^{-1}$), exceeding the system's global velocity dispersion ($\mytilde1300~\mbox{km}~\mbox{s}^{-1}$). This alone suggests that the high-speed bulk motion along the LOS direction might be an important factor to consider in our reconstruction of the merging scenario. Using the Very Large Array (VLA) telescope at 1.4 GHz, \cite{Govoni2001} confirm  the presence of the radio relic in A115, whose existence was initially hinted at by the earlier all sky radio survey \citep{Condon1998}. If we accept the belief that radio relics become detectable when the merger happens nearly in the plane of the sky, the reconciliation of the large LOS velocity with the presence of the radio relic would require an unusually large transverse velocity.

Another puzzling aspect of A115 is a large difference in the mass measurements reported in the literature \citep[e.g.,][]{Govoni2001, Barrena2007, Okabe2010, Oguri2010, Lidman2012, Sifon2015}. Although in general it is challenging to determine exact masses for merging clusters possessing complicated substructures, the A115 mass discrepancy is nearly an order of magnitude in some extreme cases. Given the potential of A115 to enhance our understanding of the plasma physics in cluster mergers, one high-priority task is to obtain the accurate mass of each substructure, as well as the global mass of the system. 
This mass information is essential when one attempts to perform a numerical simulation of the cluster merger with high accuracy.

Our multi-wavelength study of A115 has several objectives.
First, we determine the accurate mass of A115 with weak lensing (WL). Although there are several WL studies of the system in the literature, our analysis differs in several aspects.
\cite{Pedersen2007},\cite{Okabe2010}, and \cite{Oguri2010} present only a global mass of A115 without addressing the substructures.
The substructure mass estimate is a crucial input to numerical simulations. In addition, the global mass estimate itself is subject to bias when one regards the merging system as a single halo. \cite{Hoekstra2012} treat A115N and A115S separately and estimate individual masses. However, each mass estimate is obtained without subtracting the contribution from the other substructure. In general, this omission leads to overestimation of the mass.
Second, we reconstruct an accurate WL mass map and provide careful statistical analysis of the mass peak positions with respect to the ICM and optical luminosity peaks. Among the previous WL studies of A115, only \cite{Okabe2010} present a WL mass map. Interestingly, their mass peaks possess large offsets with respect to the corresponding brightest cluster galaxies (BCGs). However, since no remark on the centroid uncertainty is present, it is impossible to interpret the result quantitatively.
Third, we revisit the dynamical analysis of A115 with our new spectroscopic catalog. Because our new catalog (266) contains more than a factor of 3 times the spectroscopic cluster members of the one (88) used by \cite{Barrena2007}, the overall gain in statistical power is substantial. In particular, we re-examine the large LOS velocity difference between A115N and A115S claimed by \cite{Barrena2007}. We also compare cluster mass estimates based on improved velocity dispersion measurements.
Fourth, we provide mass estimates using deep (360 ks) {\it Chandra} data. Early {\it Chandra} studies are mostly based on relatively short exposure data. The latest study \citep{Hallman2018} utilized all existing {\it Chandra} data to provide a high-quality temperature map. However, the study did not present a representative temperature measurement for each X-ray peak and no mass estimate was given.
Finally, we present a new merging scenario of A115 consistent with our multi-wavelength data.

Our paper is structured as follows. \S\ref{sec_observation} describes our data and reduction. We explain our WL analysis in \S\ref{sec_WL_analysis}. \S\ref{sec_result} presents WL results, mass estimates from X-ray and cluster member spectroscopic data, and mass-to-light ratios. In \S\ref{sec_discussion} we discuss mass discrepancies, offsets, and a possible merging scenario before we conclude in \S\ref{sec_conclusion}.

We assume a flat $\Lambda$CDM cosmology with $H_{0}=$ 70~km s$^{-1}$ Mpc$^{-1}$, $\Omega_{m}=$ 0.3, and $\Omega_{\Lambda}=$ 0.7. At the redshift of A115, $z=0.192$, the plate scale is $\mytilde3.21$ kpc $\arcsec^{-1}$. $M_{200c}$ is defined as the mass enclosed by a sphere inside which the average density equals to 200 times the critical density at the cluster redshift. We use the AB magnitude system throughout.

\section{Observation and Data Reduction} \label{sec_observation}
\subsection{Subaru/Suprime-Cam Data} \label{sec_subaru_observation}
\begin{figure*}%[h!]
        \centering
        \includegraphics[width=0.99\textwidth]{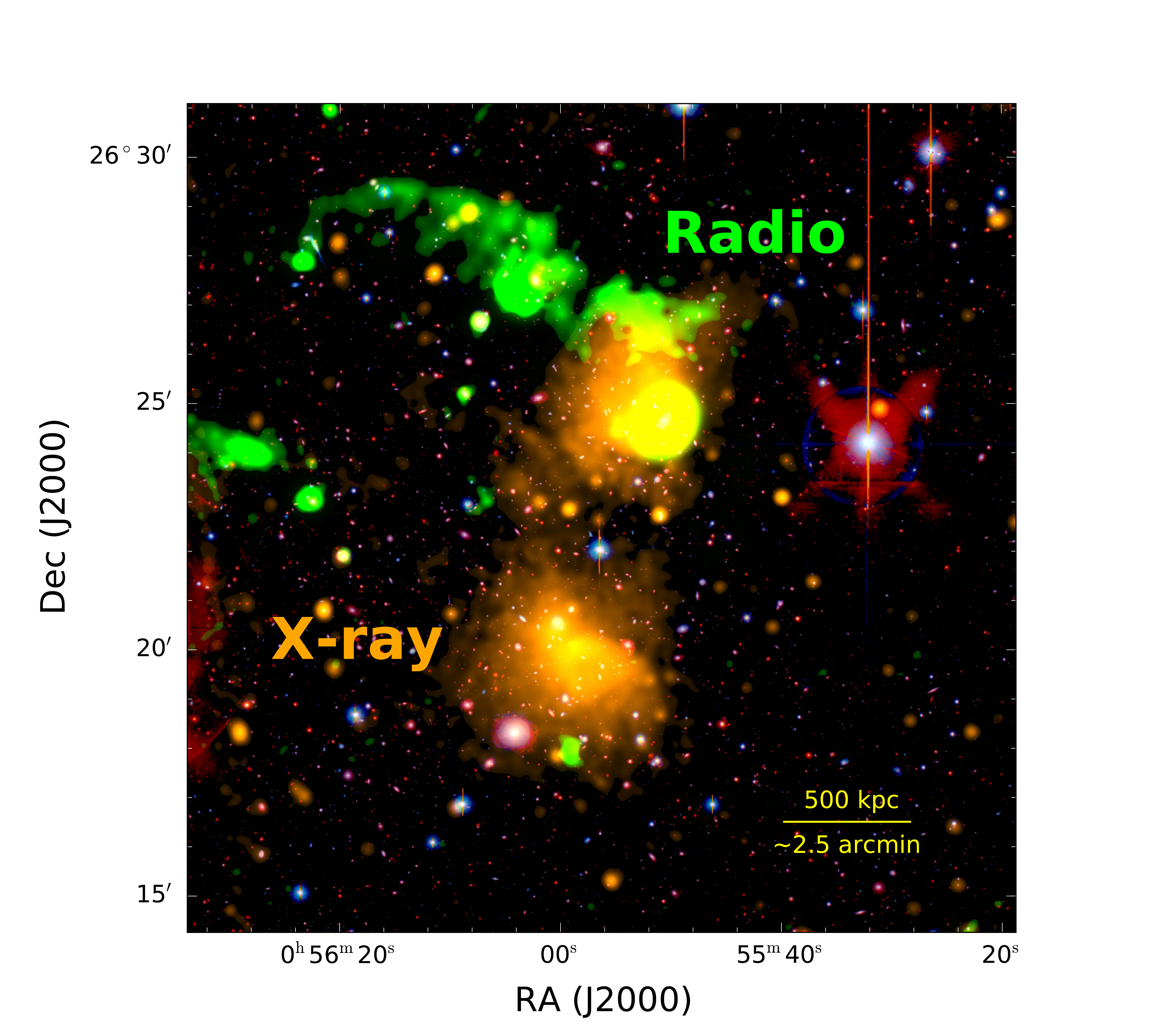}
        \caption{Color composite image of A115. Subaru/Suprime-Cam $V$, $V$+$i^{\prime}$, and $i^{\prime}$ filter images represent the intensities in blue, green, and red, respectively. Overlaid are the {\it Chandra} X-ray emission reduced in the current paper and the VLA radio images provided by \cite{Botteon2016}.
        The X-ray emission shows that A115 is comprised of two subclusters. The $\mytilde2.5$ Mpc northern radio relic stretches nearly perpendicular to the axis connecting the two X-ray emitting subclusters with the western edge terminating at the northern subcluster.
        } 
        \label{fig:color_image}
\end{figure*}

A115 was observed using the Subaru/SuprimeCam on 2003 September 25 and 2005 October 3. We retrieved the $V$- and $i^{\prime}$-band archival data from \texttt{SMOKA}\footnote{https://smoka.nao.ac.jp/}. The total integrations are 1,530~s and
2,100~s for the $V$ and $i^{\prime}$ filters, respectively. The seeings of the $V$ and $i^{\prime}$ filters are FWHM $=0.58\arcsec$ and $0.65\arcsec$, respectively. Note that the $V$-band dataset used in \cite{Okabe2010} is a subset (the total integration was 540~s) of the one used in the current study whereas their $i^{\prime}$-band dataset is identical to ours.
 
The basic CCD processing (overscan subtraction, bias correction, flat-fielding, initial geometric distortion correction, etc.) was carried out with the \texttt{SDFRED1}\footnote{https://www.subarutelescope.org/Observing/Instruments/ \\ SCam/sdfred} \citep{Yagi2002, Ouchi2004} pipeline.
We performed the rest of the imaging data reduction using our WL pipeline, which incorporates the \texttt{SCAMP}\footnote{https://www.astromatic.net/software/scamp}, \texttt{SExtractor}\footnote{https://www.astromatic.net/software/sextractor}, and \texttt{SWARP}\footnote{https://www.astromatic.net/software/swarp} packages.

We utilized the SDSS-DR9 \citep{Ahn2012} catalog to refine astrometric accuracy with \texttt{SCAMP}. A deep mosaic stack was produced
in two steps. A median mosaic image was generated with \texttt{SWARP} using the alignment information output by \texttt{SCAMP}. This median-stacking algorithm enables us to remove cosmic rays, some bleeding trails, and some CCD glitch features. However, in terms of S/N, this median-stacking result is not optimal. The final science image was created by weight-averaging individual frames, where we flagged the aforementioned, unwanted features by performing 3$\sigma$ clipping based on the median image generated in the first step.

We ran \texttt{SExtractor} in dual-image mode, which takes two images as input and uses one for detection and the other for measurement. Our detection image was created by weight-averaging the $V$- and $i^{\prime}$-band mosaic images. This dual-image mode allows us to obtain identical isophotal apertures between the two filters based on the common detection image, which is deeper than either of the two images alone.
These identical isophotal apertures are needed to obtain accurate object colors.
Photometric zeropoints were determined by using the SDSS Data Release 13 catalog that overlaps the cluster field. Because the SDSS-DR13 does not include the Johnson $V$-band, we performed a photometric transformation using the following relation \citep{Jester2005}:

\begin{equation}
    V_{\text{Johnson}}=g_{\scalebox{0.5}{SDSS}}-0.59(g_{\scalebox{0.5}{SDSS}}-r_{\scalebox{0.5}{SDSS}})-0.01.
\end{equation}
We employed isophotal magnitudes (\texttt{MAG\_ISO}) to estimate object colors, whereas total magnitude (\texttt{MAG\_AUTO}) was used to compute object luminosities.

 \subsection{{\it Chandra} Data}
We retrieved the {\it Chandra} data (ObsID: 3233, 13458, 13459, 15578, and 15581) for A115 
from the {\it Chandra} archive\footnote{http://cxc.harvard.edu/cda/}. The ObsID 3233 dataset was taken in 2002, while the other four were taken in 2012 November. All observations were carried out with the ACIS-I detector in \texttt{VFAINT} mode with 
total exposure time $\mytilde360$ ks. 
We reduced the {\it Chandra} data using the \texttt{CIAO} 4.9 pipeline and the \texttt{CALDB} 4.7.3 calibration database. The observations were re-projected to the same tangent plane and combined using the \texttt{merge$\_$obs} script. 

We created a broadband image by selecting the events within the energy range 0.5-7 keV with a $2~$pixel $\times~2$~pixel binning scheme. This broadband image was divided by our exposure map\footnote{The exposure map is an image of the effective area at each sky position and accounts for the effects of dither motion.} to produce an exposure-corrected image. In Figure~\ref{fig:color_image} this exposure-corrected image is overlayed with the VLA radio emission on our Subaru color-composite image.

In preparation for X-ray temperature measurement, we performed our initial data reduction using the \texttt{chandra$\_$repro} script. 
The \texttt{chandra$\_$repro} script automates the instrument-dependent sensitivity corrections, Charge Transfer Inefficiency (CTI) corrections, and removal of bad pixels and cosmic rays. The reduced data were reprojected to a common tangent plane using the \texttt{reproject$\_$obs} script. We masked out the point sources that are detected by the \texttt{wavdetect} script. We then constructed a lightcurve and identified background flares as detections that are 3$\sigma$ outliers. The flares were removed using the \texttt{deflare} script.

\section{Weak-Lensing Analysis} \label{sec_WL_analysis}

\subsection{Shear Measurement}
Our WL pipeline has been applied to a number of ground- and space-based imaging data \citep[e.g.,][]{Jee2013,Finner2017} and its variant has been validated in the most recent public shear testing program \citep{Mandelbaum2015}. Readers are referred to \cite{Finner2017} for details. Here we present a brief summary of our PSF model and ellipticity measurement.

\subsubsection{PSF Modeling} \label{sec_psfmodeling}

Point spread function (PSF) modeling is a crucial step in a
WL study. Unless corrected for, the PSF not only dilutes the lensing signal, but also induces a distortion mimicking WL.
In this study, we use the principal component analysis (PCA)
approach \citep{Jee2007,JeeTyson2011}. 

The observed PSF at a specific location on the mosaic is a combination of the PSFs from all contributing frames. Thus, to properly consider each component, we modeled the PSF for each contributing frame and then stacked them to a final PSF model. 

\begin{figure*}

        \centering
        \includegraphics[width=0.95\textwidth]{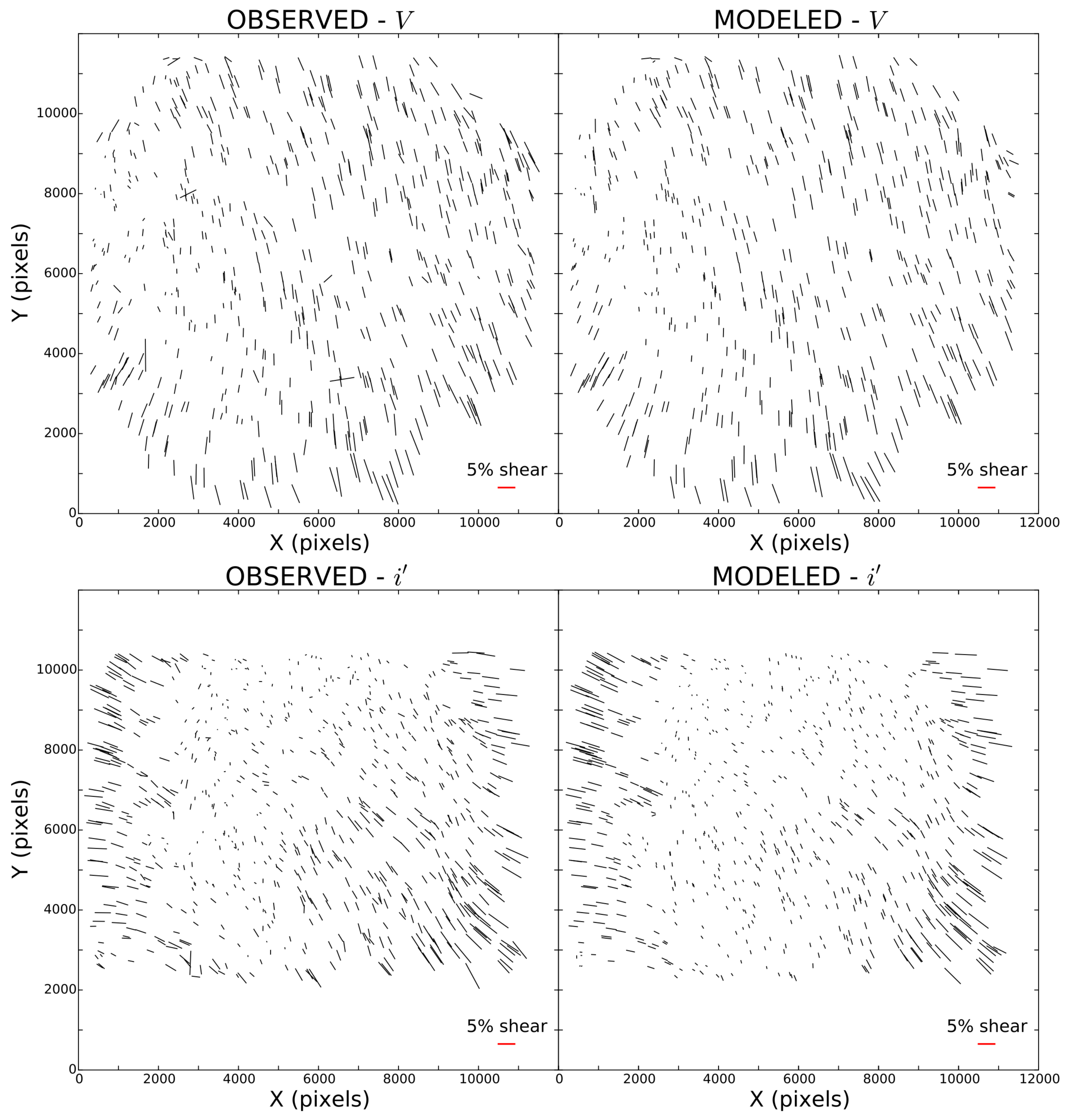}
        \caption{
        Comparison between the observed and model PSFs. The length of the stick represents the magnitude of the star/PSF ellipticity while the orientation shows the direction of elongation. The observed PSF ellipticities are measured from the star images in our coadd image. The model PSFs are created by stacking all contributing PSFs (modeled with PCA) from individual exposures.
        Top: For the $V$-filter, the position-dependent ellipticity variation of the model PSFs closely matches that of the observed stars, which indicates that our model is a robust representation of the observed PSF ($\left< \delta e^2 \right>^{1/2}\sim0.014$).
        Bottom: For the $i^{\prime}$-filter, the agreement between model and observation is not as accurate as the one for the $V$-filter ($\left< \delta e^2 \right>^{1/2}\sim0.027$).
        } 
        \label{fig:PSF_com}
\end{figure*}

One way to examine the fidelity of the PSF model is to compare the ellipticity pattern of the mosaic fields between observation and model as shown in Figure \ref{fig:PSF_com}. The left panel shows the ellipticity pattern of the observed stars and the right panel shows the pattern reconstructed by our PSF model. For the $V$ filter (top), both  magnitude and direction of the PSFs across the mosaic field are closely reproduced. The mean residual rms is $\left< \delta e^2 \right>^{1/2}\sim0.014$ per ellipticity component. 
The good agreement demonstrates that the PCA-based PSF model is robust. Also, it demonstrates that the image co-adding alignment is performed with high fidelity; even a subpixel-level misalignment would manifest itself as a noticeable PSF ellipticity pattern in the co-add image (left panel), which however could not be reproduced by the model (right panel) that assumes a perfect alignment. For the $i^{\prime}$ filter (bottom), we could not make the model PSF ellipticity pattern match the observed pattern as accurately as in the case of the $V$ filter. The mean residual rms in this case is $\left< \delta e^2 \right>^{1/2}\sim0.027$, which is nearly a factor of two larger. Currently, the exact source of this poor match between model and observation is unknown.

We decide to measure WL signals from our $V$-band image, for which our PSF model is more accurate. An additional merit from using the $V$-band data rather than the $i^{\prime}$-filter is its smaller PSF ($\mytilde11$\% smaller on average). Given the same PSF model accuracy, smaller PSFs provide more reliable shapes for fainter and smaller galaxies, which have higher chances of being background and thus dominate WL signals.

\subsubsection{Ellipticity Measurement} \label{sec_ellipticity}

We fit a PSF-convolved elliptical Gaussian to a galaxy image to determine its two ellipticity components $e_1$ and $e_2$, which we define as
\begin{equation}
\begin{aligned}
    e_1 & = & e \cos 2\theta, \\
    e_2 & = & e \sin 2\theta, \\
    e & = & \frac{a-b}{a+b}
\end{aligned}
\end{equation}
\noindent
where $a$ and $b$ are the semi-major and semi-minor axes of the best-fit elliptical Gaussian, respectively, and $\theta$ is the position angle of the semi-major axis. Since the elliptical Gaussian is convolved with a model PSF when fitted to the galaxy image, the resulting ellipticity is corrected for PSF systematics.

The elliptical Gaussian profile contains seven free parameters: normalization, two parameters for centroid, semi-major axis, semi-minor axis, position angle, and background level. We fixed the centroid and background level using the \texttt{SExtractor} outputs \texttt{X\_IMAGE}, \texttt{Y\_IMAGE}, and \texttt{BACKGROUND}, respectively. This reduces the number of free parameters to four, which improves convergence for faint sources. We used the $\chi^2$ minimization code \texttt{MPFIT}\footnote{https://www.physics.wisc.edu/\mytilde craigm/idl/fitting.html} to fit the model to the galaxy image and estimate the ellipticity uncertainty. 

In general, this raw ellipticity is a biased measure of the true shear for a number of reasons \citep[e.g.,][]{Mandelbaum2015}. The bias is often expressed as $\gamma = (1+m_{\gamma}) e  + m_{\beta}$, where $m_{\gamma}$ and $m_{\beta}$ are often referred to as ``multiplicative" and ``additive" biases, respectively.
We find that although the additive bias is negligible for our WL pipeline, the multiplicative bias is not \citep{Jee2013}. From our image simulation, we determine $m_{\gamma}=0.15$ for our source population. This multiplicative factor is applied to our ellipticity catalog.

\subsection{Source Selection}
\label{sec_source_selection}
Only light from galaxies located at a greater distance than the cluster is lensed by the gravitational potential of the cluster. Ideally, one can use a photometric redshift technique to enable efficient selection of background galaxies. However, this is not feasible in our case, where only two broadband filters are available. Therefore, in the current study we used a color-magnitude relation to select source galaxies. 

\begin{figure}[h!]
        \centering
        \includegraphics[width=0.46\textwidth]{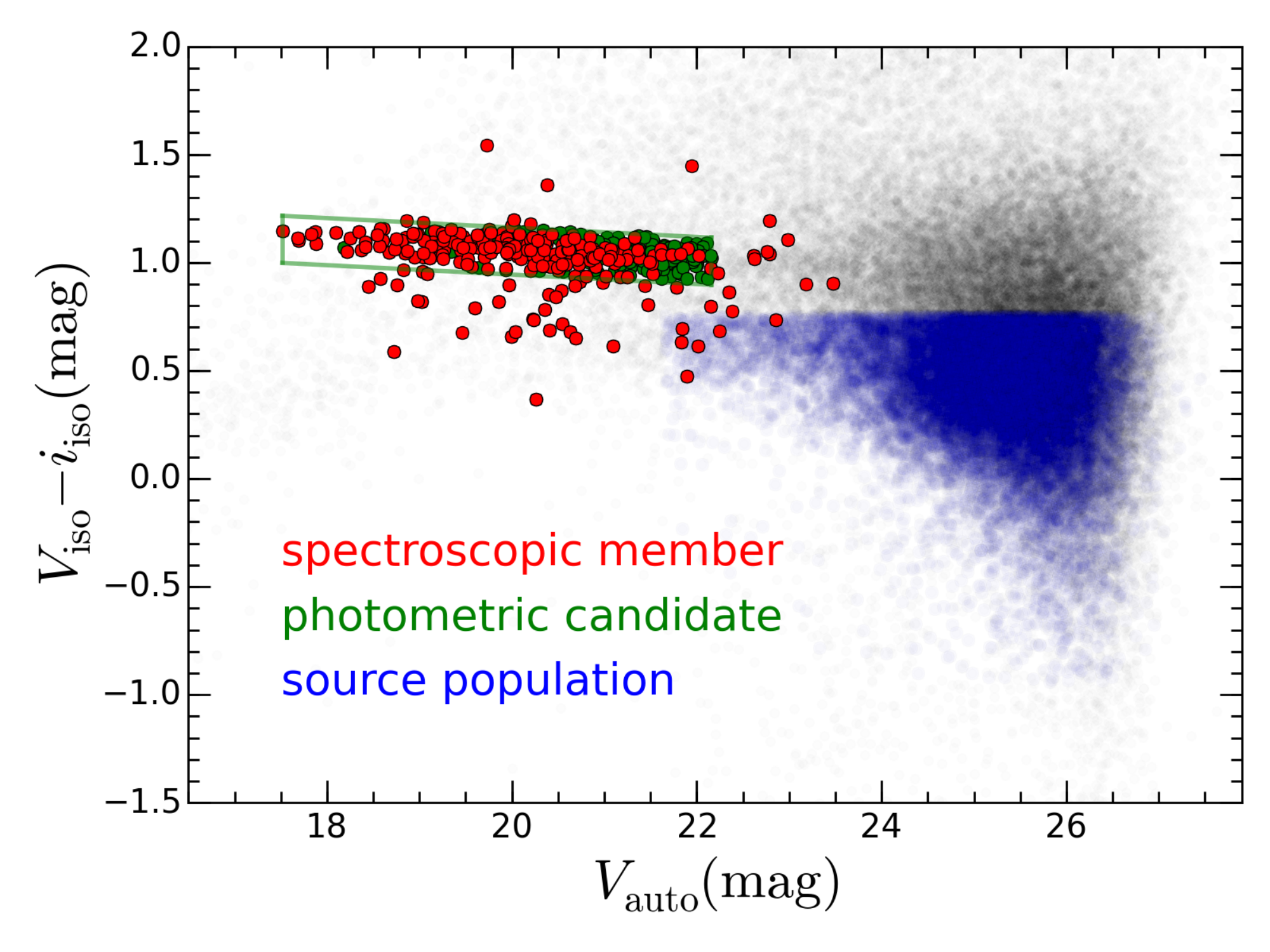}
        \caption{
        Color-magnitude relation in the A115 field. 
        Galactic dust reddening has been corrected for using \cite{Schlegel1998}. Red-sequence galaxies show a tight color-magnitude relation. Red circles are spectroscopically confirmed cluster members and green circles are photometric member candidates based on the color-magnitude relation and our visual inspection of the galaxy morphology of each object. The green parallelogram depicts the color and magnitude selection criteria for the selection of photometric member candidates.
        Blue circles are the galaxies that populate our source catalog, as selected by the criteria in Table 1. Both spectroscopic members and photometric candidates are utilized to estimate the number and luminosity density of the cluster. Only spectroscopic members are used for the dynamical mass estimation.
        }
        \label{fig:CMD}
  
\end{figure}

Figure \ref{fig:CMD} shows the color-magnitude diagram (CMD) of the A115 field. It is clear that a majority of the early-type galaxies of A115 show a tight color-magnitude relation. We selected galaxies that are bluer and fainter than this red-sequence to minimize the contamination of our source catalog by cluster galaxies. This selection scheme is based on the general trend that more distant galaxies are bluer and fainter than the cluster red sequence at $z\sim0.2$. 
Obviously, this trend is only roughly true and thus some fraction of the sources defined in this way are not behind the cluster. We estimated this fraction in our source redshift estimation (\S\ref{sec_redshift_estimation}).

We further refined our source catalog by imposing size and ellipticity error conditions. Objects whose semi-minor axis $b$ is smaller than 0.3 pixels were discarded because they are usually indistinguishable from stars. We require that the ellipticity error is below 0.25. This removes not only low S/N objects, but also point sources, which tend to have large ellipticity errors (in principle, stars should have no shape after PSF deconvolution). Many spurious sources are removed by the above ellipticity error and size conditions. As a further measure, we discarded sources whose ellipticities are greater than 0.9 because they are in general too elongated to be a galaxy. The last selection criteria that we applied is an \texttt{MPFIT STATUS} = 1 (a good fit).

After all selection criteria were applied, some spurious objects still survived. These objects mostly appear on  diffraction spikes and reflection rings from bright stars. We removed the spurious objects by visual inspection.
These spurious features are particularly important near A115N where a bright star with diffraction spikes is located $\mytilde4\arcmin$ west. Our final source catalog has $\mytilde$17,000 galaxies over the $\mytilde600$ arcmin$^2$ area. The resulting source density $\mytilde24$ arcmin$^{-2}$ is a factor of two larger than the one used in \cite{Okabe2010}.
We summarize our source selection criteria in Table \ref{tab:source_criteria}.
\begin{table}[h]
\centering
\caption{Source Selection Criteria}
\begin{tabular}{p{5cm}p{2.8cm}}
\hline \hline
%\multicolumn{2}{c} {Source Selection Criteria} \\
%\hline
Magnitude            & $21.5<V<27.5$                                   \\
Color index          & $-1<V-i<0.7$                                    \\
Ellipticity          & $e<0.9$                                         \\
Ellipticity error    & $\sigma_{e}<0.25$                                    \\
Semi-major axis      & $a<30$                                          \\
Semi-minor axis      & $b>0.3$                                         \\
SExtractor Flag                 & $f<4$                                           \\
MPFIT status         & $s=1$\\
\hline \hline
\end{tabular}
\label{tab:source_criteria}
\end{table}

\subsection{Redshift Estimation of Source Population}
\label{sec_redshift_estimation}
Quantitative interpretation of a lensing signal requires information on the redshift distribution of the source population. The observed shears that are extracted from the source galaxies are expressed in units of the critical surface density $\Sigma_c$ defined as 
\begin{equation}
    \Sigma_c = \frac{c^2}{4 \pi G D_l \beta},
\end{equation}
\noindent
where $c$ is the speed of light, $G$ is the gravitational constant, $D_l$ is the angular diameter distance of the lens, and $\beta$ is the lensing efficiency.
The lensing efficiency is given by
\begin{equation}\label{eq:lensing_efficiency}
    \beta = \left \langle \mbox{max}\left ( 0,\frac{D_{ls}}{D_s} \right ) \right \rangle,
\end{equation}
\noindent
where $D_{s}$ and $D_{ls}$ are the angular diameter distances to the cluster and from cluster to source galaxy, respectively. 
Note that objects with negative $\beta$ values are assigned a zero value because foreground sources do not contribute to the lensing signal regardless of their redshifts.
%%%%%%%%%%%%%%%%
Since we do not have photometric redshifts for individual galaxies, we evaluated $\beta$ for the source population statistically using a control field. This requires the assumption that the statistical properties of the control field are similar to those of the A115 field. One may be concerned that this assumption might be invalid when we compare two small fields because of the sample variance. \cite{Jee2014} investigated the issue in their mass estimation of the galaxy cluster ACT-CL~J0102−4915. They found that even for their $6\arcmin\times 6\arcmin$ field the effect of the sample variance is small, responsible for only $\mytilde4$\% shift in mass. This is mainly because the image is deep and thus produces a large redshift baseline for the source galaxy distribution.
In the current study, where the field is much larger with a comparable depth, we expect that the sample variance is also sub-dominant.

\begin{figure}[h!]
        \centering
        \includegraphics[width=0.46\textwidth]{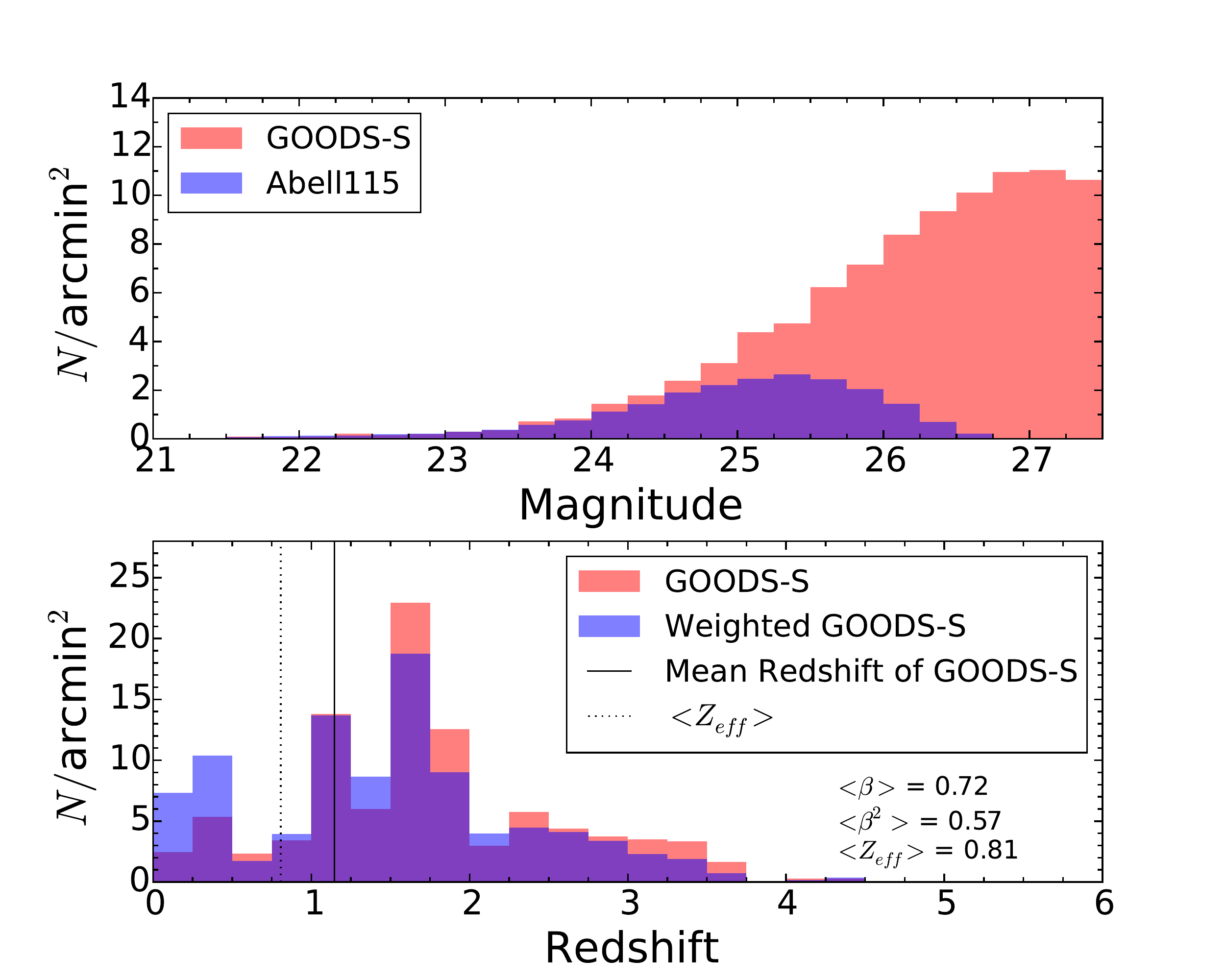}
        \caption{Estimation of the effective redshift for our source population. We utilized the GOODS-S photometric redshift catalog \citep{Dahlen2010} as our control field. The top panel compares the magnitude distributions between our sources and the GOODS-S galaxies after application of the same source selection criteria. The density of the GOODS-S galaxies is higher because of the difference in depth and de-blending resolution. The GOODS-S galaxies were weighted by the ratio of our source density to the control field density when we estimate the effective redshift of our source population.
        The bottom panel shows the resulting redshift distribution before (blue) and after (red) this weighting. 
        }
        \label{fig:redshift_est} 

\end{figure}

We chose the Great Observations Origins Deep Survey South \citep[GOODS-S;][]{Giavalisco2004} data as our control field and utilized the photometric redshift catalog of \cite{Dahlen2010}. After applying the same color and magnitude selection criteria (Table \ref{tab:source_criteria}) on the GOODS-S catalog, we compared its magnitude distribution (red bins) with that in the source population (blue bins), as shown in the top panel of Figure~\ref{fig:redshift_est}. Since the GOODS-S images are deeper, 
and its galaxies are better de-blended, the number density of objects per magnitude bin is much higher in GOODS-S at $i^{\prime}\gtrsim25$.
To account for this difference, we weighted the redshift distribution of the GOODS-S catalog for each magnitude bin by the number density ratio of our source catalog to the GOODS-S catalog (see the bottom panel of Figure~\ref{fig:redshift_est}). With the GOODS-S conformed to our source catalog, we measured the average lensing efficiency by Equation \ref{eq:lensing_efficiency}. The lensing efficiency obtained in this way is $\beta=0.72$, which corresponds to the effective redshift $z_{\rm eff}=0.81$. This $\beta$ value is similar to the estimate $\beta=0.701$ reported in \cite{Okabe2010}, whose source shape measurement is based on a 1500s $i^{\prime}$-band image.
The assumption that all sources are located at this single redshift causes bias in cluster mass estimation \citep[as discussed in][]{Seitz1997, Hoekstra2000}. To correct for this bias, we applied the following correction to the observed shear:
\begin{equation} \label{eq:observed_shear}
\begin{split}
    g' & =  \left [ 1+\left ( \frac{\left \langle \beta^2 \right \rangle}{ \left \langle \beta \right \rangle^2} -1 \right ) \kappa \right ] g \\
    & =  \left ( 1+0.10 \kappa \right )g,
\end{split}
\end{equation}
where $\langle \beta^2 \rangle \sim 0.57$.

\begin{figure}[h!]
        \centering
        \includegraphics[width=0.46\textwidth]{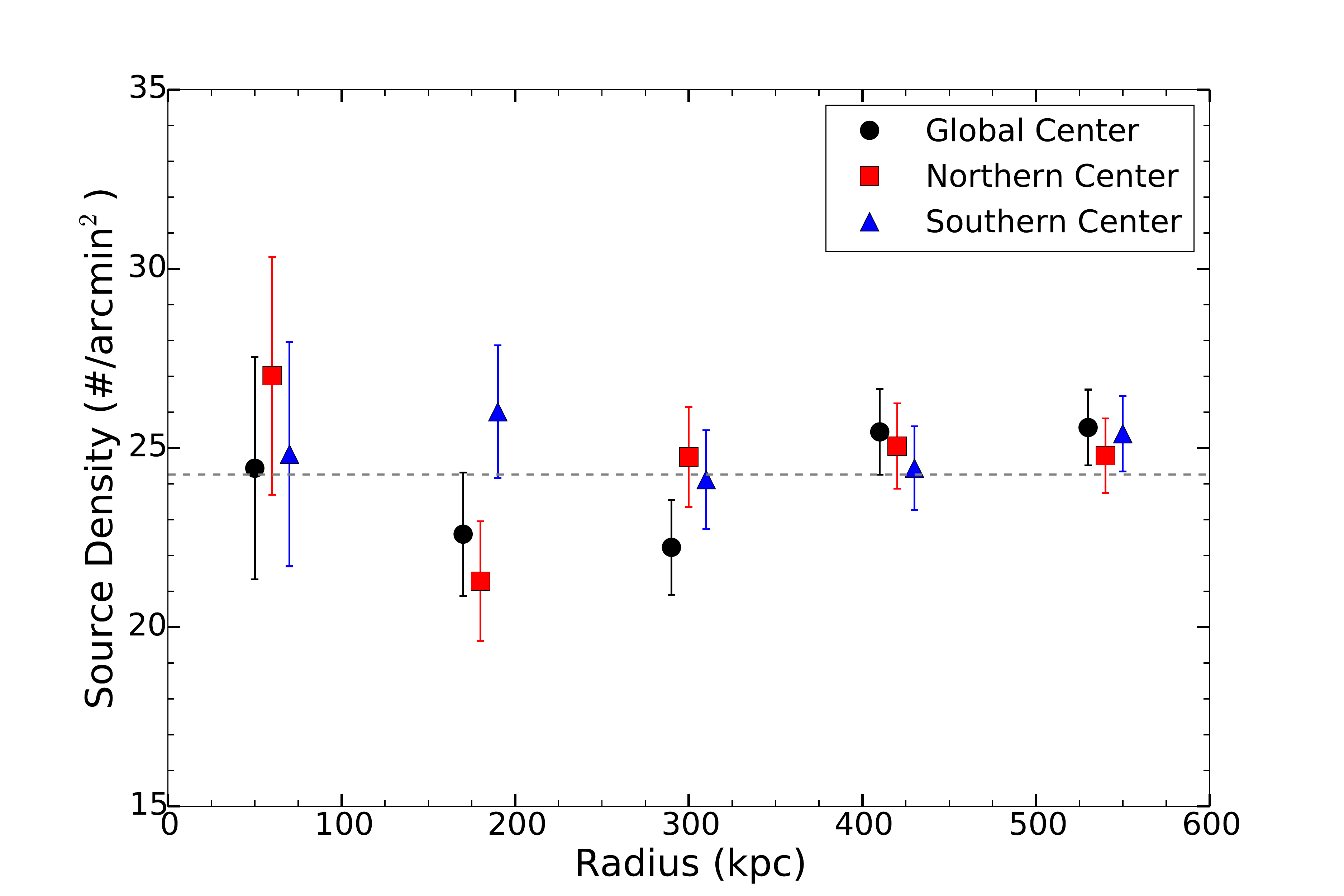}
        \caption{Source density profile as a function of projected distance from each cluster center. The profiles are centered on the global center (black), northern cluster (blue), and southern cluster (red). Dashed line indicates the mean source density of the cluster field. No significant excess at small radii is found.
        }
        \label{fig:source_density} 
\end{figure}

Another concern in this procedure might be blue cluster member contamination. Given the current limited filter coverage, it is difficult to efficiently select and remove the blue cluster members. If the contamination is significant, this will lead to underestimation of the lensing signal (thus underestimation of the cluster mass). However, our previous studies found that the contamination is insignificant when sources are selected based on the color-magnitude relation as done in the current study. For example, in their {\it Hubble Space Telescope} WL analysis, \cite{Jee2014} compared the magnitude distribution of the sources in A520 at $z\simeq0.2$ with those in their control fields. If the blue member contamination is significant, the source density should show an excess with respect to those in the control fields. However, no such excess was found in their study. Since the redshift of A520 is comparable to that of A115, we argue that the conclusion of \cite{Jee2014} is applicable to the current study. Note that we could not perform a similar analysis with the current Subaru imaging data because of the large difference in instrument resolution between the cluster and control fields. 
Instead, we examined the source density as a function of cluster-centric distance. If the blue member contamination is significant, the source density profile may show a peak near the cluster center. Figure \ref{fig:source_density} shows that for all three choices of centers, the source densities at small radii have no significant excess.

\section{Results}
\label{sec_result}

\subsection{Mass Reconstruction} \label{section:mass reconstruction}

\begin{figure}[h!]
        \centering
        \includegraphics[width=0.5\textwidth, trim=0cm 0cm 1cm 1cm]{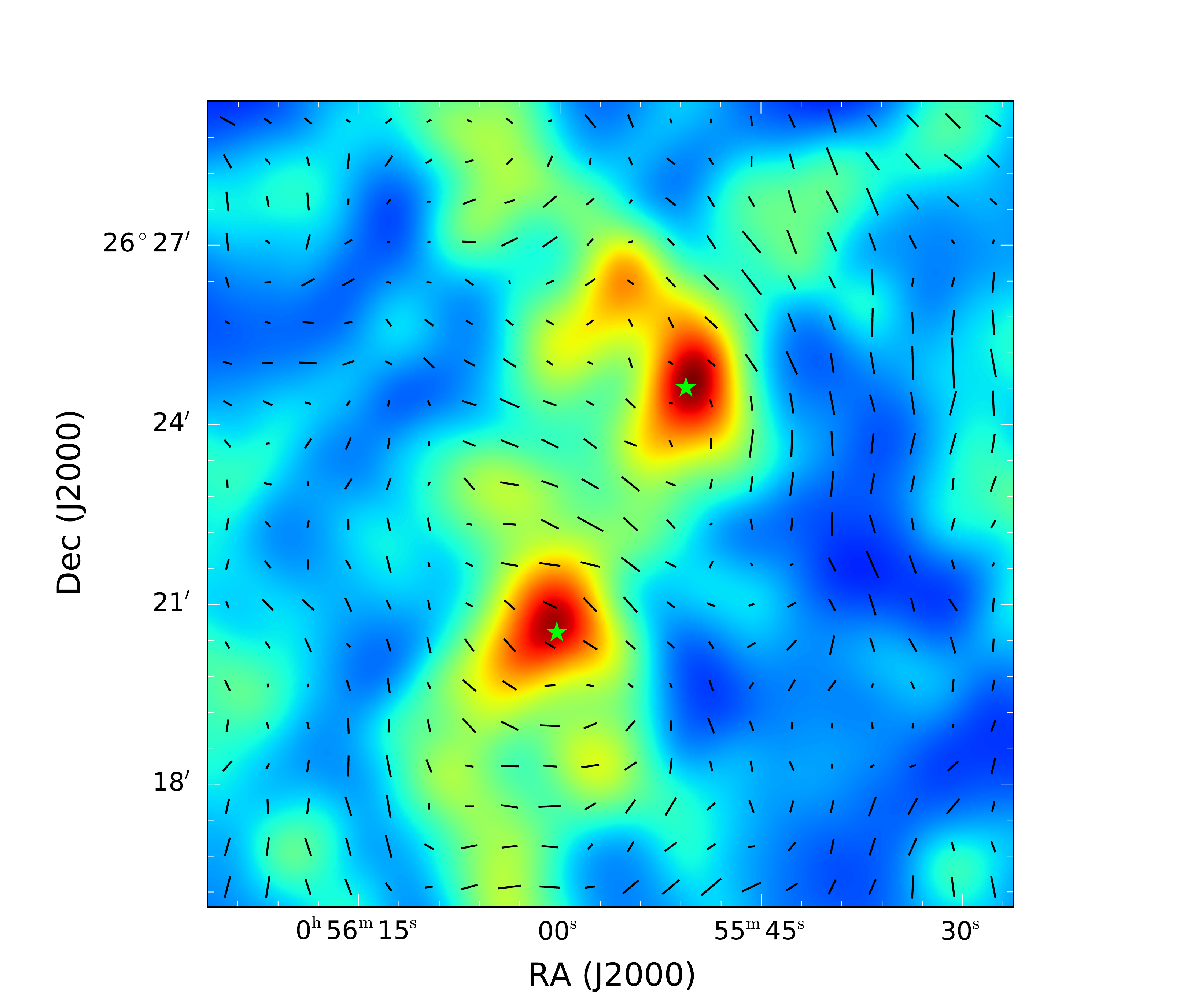}
        \caption{``Whisker" plot over convergence map. Each whisker is the reduced shear determined by averaging over the background galaxy ellipticity within an $r=80"$ circle. Green star markers indicate the position of each BCG. The length and orientation of each whisker indicate the magnitude and direction of the reduced shear, respectively. The reduced shear tends to be tangentially aligned around the mass peak and decreases with the distance from the mass center. The convergence (color-coded) was reconstructed using the maximum-entropy-maximum-likelihood method \citep{Jee2007b}. The mass map clearly reveals the bimodal structure of A115.
        } 
        \label{fig:whisker}
\end{figure}

The shapes of lensed  galaxy images are sheared by a small amount, which is typically a tiny fraction of the intrinsic shape noise.
Thus, measurement of these shears requires averaging over a large sample of background galaxies. The ``whisker plot'' in Figure \ref{fig:whisker} shows the shear in the A115 field obtained by averaging over the background galaxy ellipticities. Each whisker in the 20$\times$20 grid represents the magnitude and direction of the local average ellipticity within a radius of $r=80\arcsec$. 

The shear $\gamma$ can be converted to the surface mass density $\kappa$ (convergence) map using the following relation \citep[][hereafter KS93]{Kaiser1993}:
\begin{equation}
\kappa (\bvec{x}) = \frac{1}{\pi} \int D^*(\bvec{x}-\bvec{x}^\prime) \gamma (\bvec{x}^\prime) d^2 \bvec{x}, \label{k_of_gamma}
\end{equation}
\noindent
where $D(\bvec{x} ) = - 1/ (x_1 - i x_2 )^2$ is the transformation kernel. A number of algorithms exist for this $\gamma$-to-$\kappa$ conversion in the literature. 

\begin{figure*}
        \centering
        \includegraphics[width=0.99\textwidth]{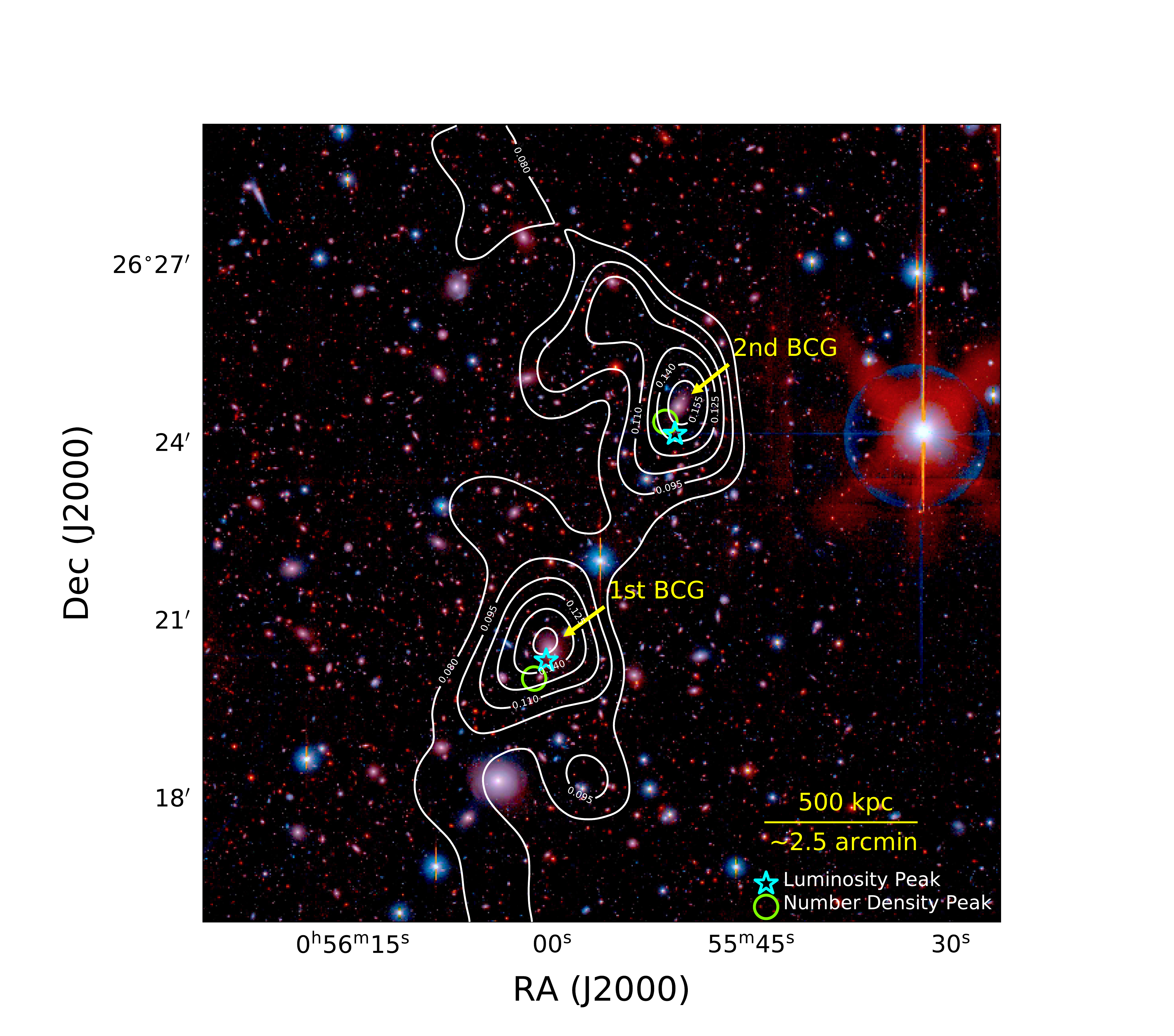}
        \caption{
        Mass reconstruction over color composite. The northern and southern mass clumps are detected at a significance of 3.8~$\sigma$ and 3.6~$\sigma$, respectively. The two mass centroids are in excellent agreement with the locations of the two BCGs.}
        \label{fig:convergence_map}
\end{figure*}

\begin{figure*}
        \centering
        \includegraphics[width=0.99\textwidth]{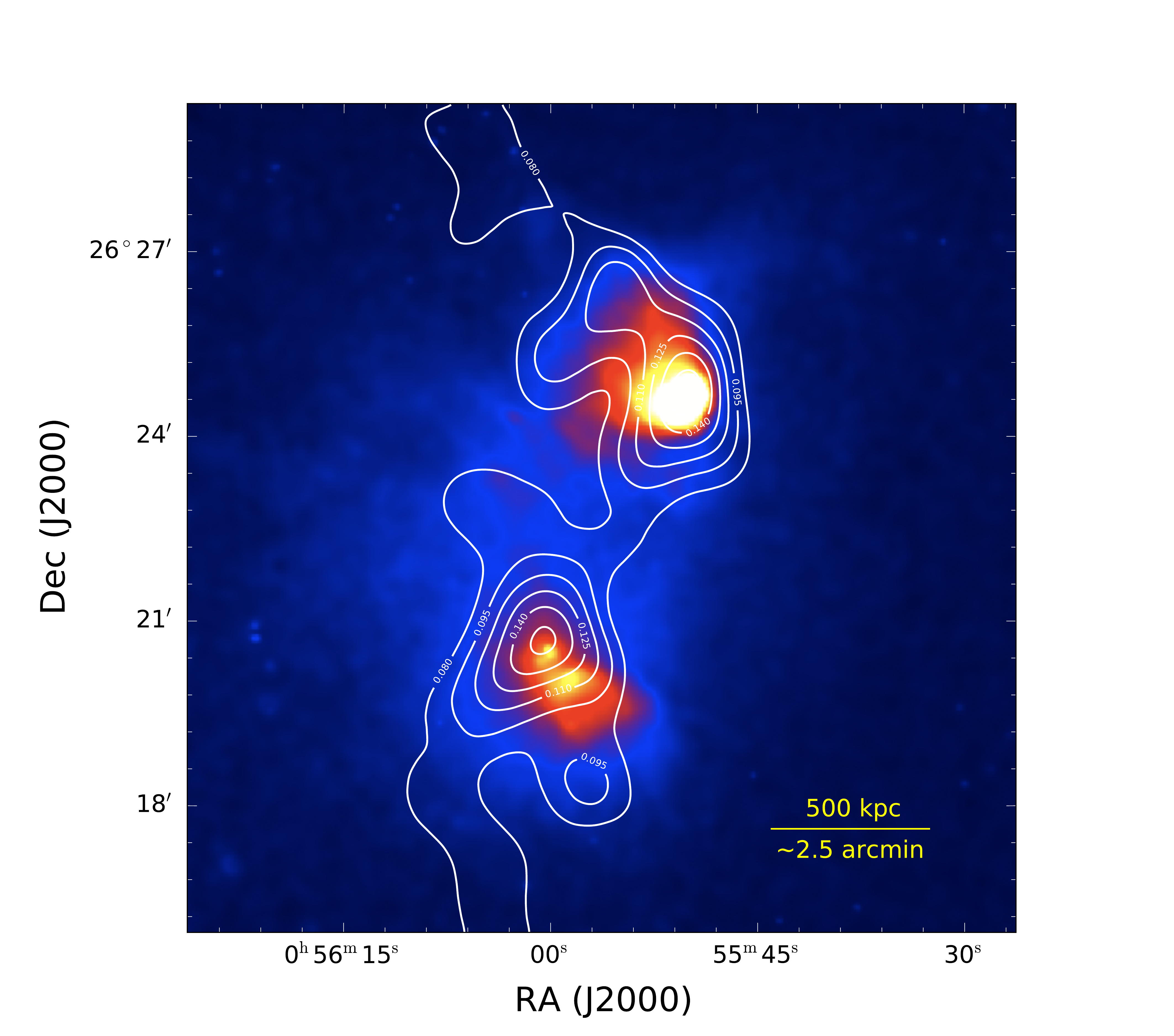}
        \caption{Exposure-corrected {\it Chandra} X-ray image overlaid with convergence contours. The image was adaptively smoothed and point sources are removed. Each mass peak agrees nicely with the corresponding X-ray peak. 
        } 
        \label{fig:x_ray_mass_dist}
\end{figure*}

In this study, we used the maximum entropy maximum likelihood method (MAXENT) described in \cite{Jee2007b} for our mass reconstruction. The MAXENT method
utilizes the ``entropy" of the pixels to regularize the mass map. This enables us to reveal high-resolution features where the S/N is high while it reduces the noise by applying large smoothing kernels in the low S/N region such as field boundaries.
Color-coded in Figure \ref{fig:whisker} is the resulting $\kappa$ map, which presents two prominent mass peaks. When we used the traditional KS93 inversion method, we recovered similar features near the mass peaks with a FWHM$\sim50\arcsec$ Gaussian kernel. 
The $\kappa$ contours are overlayed on the Subaru color-composite image in Figure \ref{fig:convergence_map}, where we see an excellent spatial agreement between both the two BCGs and the mass peaks ($\lesssim10\arcsec$, 32 kpc); the two mass peaks also coincide with the two X-ray peaks (Figure~\ref{fig:x_ray_mass_dist}).
Our bootstrapping analysis based on the KS93 reconstruction (see \S\ref{sec_offset}) shows that the northern and southern mass clumps are detected at a significance of 3.8~$\sigma$ and 3.6~$\sigma$, respectively, and the two mass centroids are highly consistent with the BCGs.

\subsection{Weak-lensing Mass Estimation}
\label{sec_wl_mass_estimation}
Many WL studies estimate galaxy cluster masses, based on the assumption that they are comprised of a single halo. However, this assumption can lead to non-negligible mass bias if substructures' masses are comparable, as in the case of A115. In this study, our main results were obtained by simultaneously fitting two Navarro Frenk-White \citep[NFW;][]{Navarro1997} halo profiles to A115N and A115S. The NFW shear model derived in \cite{Wright2000} was adopted. However, we also present the results from one-dimensional (azimuthally averaged) profile fitting centered on each substructure for comparison. This enables us to assess the amount of bias that would have been introduced if only the one-halo fitting method had been used. Also, the comparison provides a sanity check for the two halo fitting method, which is numerically less stable and requires more complicated procedure.

%In this study, our main results are obtained by simultaneously fitting two Navarro-Frenk-White \citep[NFW,][]{Navarro1997} halo profiles to A115N and A115S. However, we also present the results from one-dimensional (azimuthally averaged) profile fitting centered on each substructure for consistency check.

\subsubsection{One-dimensional Profile Fitting}
\label{sec_1dfitting}
\begin{figure*}[!htb]
        \centering
        \includegraphics[width=0.48\textwidth,trim=3cm 0cm 3cm 0cm]{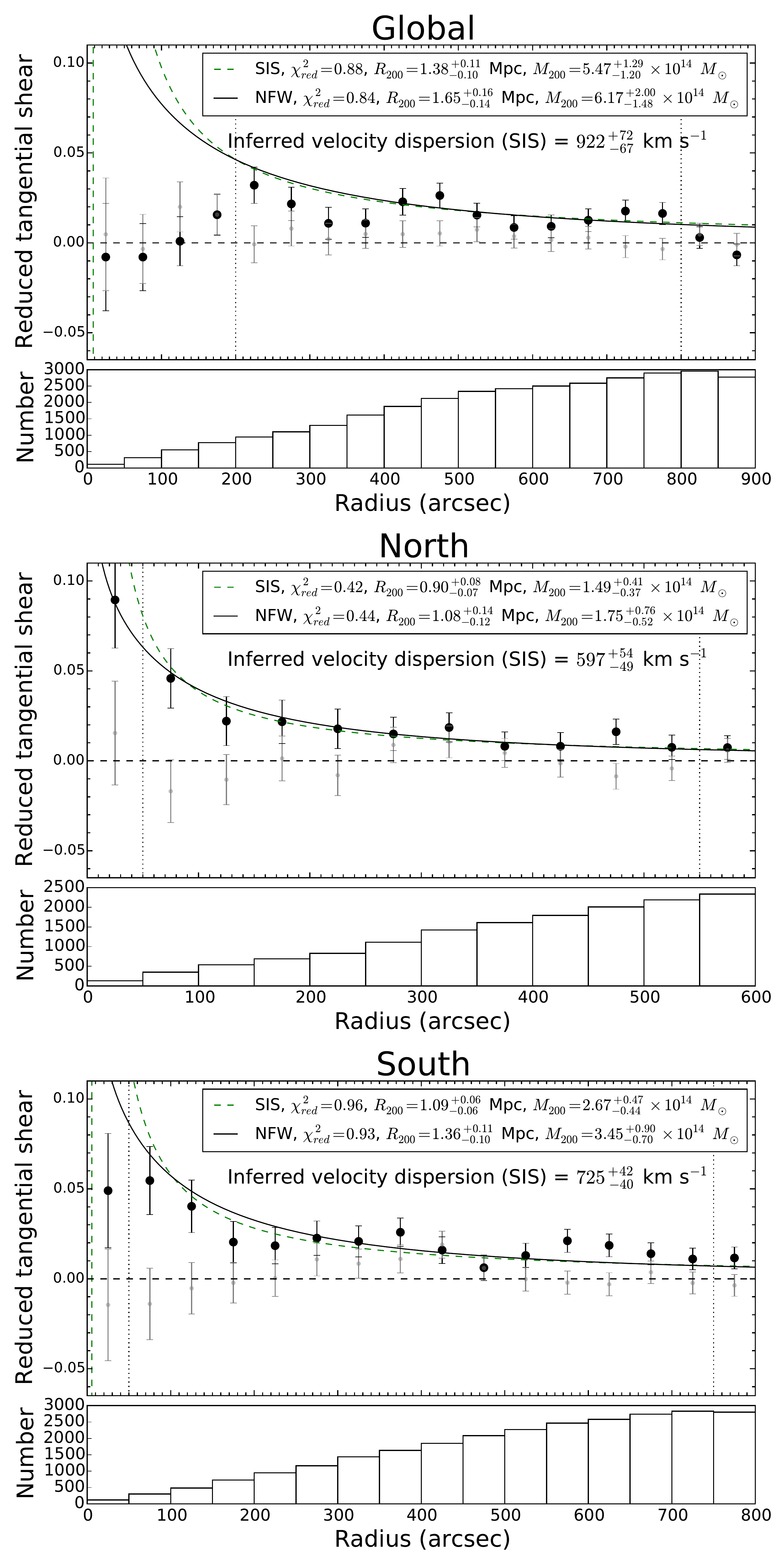}
        \caption{
        Reduced tangential shear profiles of A115. Black circles represent the azimuthally averaged tangential shear in each annulus. Gray circles show the cross shear and should be consistent with zero, as shown, when WL systematics are negligible. Solid and dashed lines indicate the best-fit NFW and SIS models, respectively. Vertical dotted lines denote the minimum and maximum radii between which we include the data for fitting. 
        See text for descriptions on our center choices and criteria for the minimum and maximum radii.}
        \label{fig:1D_mass}
\end{figure*}

The first step in one-dimensional profile fitting is the construction of the azimuthally-averaged tangential shear profile as a function of radius. Tangential shear is defined as
\begin{equation}
 g_T  = -  g_1 \cos 2\phi - g_2 \sin 2\phi \label{tan_shear},
\end{equation}
\noindent
where $\phi$ is the position angle of the source with respect to the subcluster center, and $g_1$ and $g_2$ are the two components of the calibrated ellipticity. Figure \ref{fig:1D_mass} shows the three tangential shear profiles when the center is placed at the global, A115N, and A115S centers. For the two subclusters, we chose the BCG locations as their centers since the centroids of the three cluster constituents (BCG, X-ray emission, and WL mass) agree nicely. We adopted the mean of the two subcluster peaks as the global center. If our weak-lensing resolution had been  poorer (e.g., if the number density of sources had been less than 10 $\mbox{arcmin}^{-2}$), we would have detected only a single mass clump centered near this middle point.

In Figure \ref{fig:1D_mass}, weak-lensing signals are clearly detected in all three cases nearly out to the field boundary ($r\sim900\arcsec$).
The consistency of the cross shears (obtained by rotating the position angle by 45$\degr$) with zero indicates that no significant B-modes are present in our analysis.

It is a common practice to discard signals at small radii in model fitting because of a number of issues. First, the weak-lensing assumption is violated near the cluster center. Since galaxy images are sheared non-linearly, the measurement performed without any correction can lead to cluster mass bias. Second, cluster member contamination is highest near the center, which can suppress the lensing signal. Third, the shape of the profile at small radii is sensitive to the choice of the center and the true center is unknown. Fourth, we expect baryonic effects to be non-negligible in the central region, which can make the actual profile differ from the NFW one. Currently, no consensus exists for the choice of a cuttoff radius except that it should increase with halo mass. We chose $r_{\text{cut}}=50\arcsec$ when the center was placed on each subcluster while this threshold was increased to $r_{\text{cut}}=200\arcsec$ for the global mass estimation. This increase is needed to reduce the impact of the cluster substructures on the tangential shear profile; the projected distance from the global center to a subcluster is $\mytilde150\arcsec$.
We also exclude the tangential shears at large radii if the measurements come from incomplete annuli.

We used the mass-concentration relation from \cite{Dutton2014} to characterize our NFW halo. From our one-dimensional NFW fitting, we determine the masses of A115N and A115S to be $M_{200c}=1.75_{-0.52}^{+0.76}\times10^{14}M_{\sun}$ and $3.45_{-0.70}^{+0.90}\times10^{14}M_{\sun}$, respectively. The global mass is estimated to be $6.17_{-1.48}^{+2.00}\times10^{14}M_{\sun}$ (Table~\ref{tab:1D_tangential_shear_NFW}). Consistent masses are obtained when we assume a singular isothermal sphere (SIS) instead (Table~\ref{tab:1D_tangential_shear_SIS}). We used these SIS fitting results to evaluate inferred velocity dispersions. The reduced $\chi^2$ values show that both models describe the observed profiles reasonably well and there is no significant indication that one model is preferred over the other. 

\begin{table}[]
\centering
\caption{1D NFW Profile Fitting Result}
\label{tab:1D_tangential_shear_NFW}
\begin{tabular}{cccc}
\hline \hline
       & \begin{tabular}[c]{@{}c@{}}$R_{200c}$\\ (Mpc)\end{tabular} & \begin{tabular}[c]{@{}c@{}}$M_{200c}$\\ $(\times 10^{14} M_{\sun})$\end{tabular} & \begin{tabular}[c]{@{}c@{}}$\chi^{2}_{red}$\end{tabular} \\
       \hline
Global & $1.65^{+0.16}_{-0.14}$    & $6.17^{+2.00}_{-1.48}$    & 0.84  \\
North  & $1.08^{+0.14}_{-0.12}$    & $1.75^{+0.76}_{-0.52}$    & 0.44  \\
South  & $1.36^{+0.11}_{-0.10}$    & $3.45^{+0.90}_{-0.70}$    & 0.93 \\ \hline \hline
\end{tabular}
\end{table}

\begin{table}[]
\centering
\caption{1D SIS Profile Fitting Result}
\label{tab:1D_tangential_shear_SIS}
\begin{tabular}{ccccc}
\hline \hline
       & \begin{tabular}[c]{@{}c@{}}$\sigma_v$\\ (km s$^{-1}$)\end{tabular} & \begin{tabular}[c]{@{}c@{}}$R_{200c}$\\ (Mpc)\end{tabular} & \begin{tabular}[c]{@{}c@{}}$M_{200c}$\\ $(\times10^{14} M_{\sun})$\end{tabular} & $\chi^{2}_{red}$ \\ \hline
Global & $922^{+72}_{-67}$    & $1.38^{+0.11}_{-0.10}$    & $5.47^{+1.29}_{-1.20}$  & 0.88  \\
North  & $597^{+54}_{-49}$    & $0.90^{+0.08}_{-0.07}$     & $1.49^{+0.41}_{-0.37}$  & 0.42  \\
South  & $725^{+42}_{-40}$    & $1.09^{+0.06}_{-0.06}$    & $2.67^{+0.47}_{-0.44}$  & 0.96 \\ \hline
\end{tabular}
\end{table}

\subsubsection{Two-dimensional Simultaneous Profile Fitting with Two Halos} \label{sec:two_halo_fitting}

The results presented in \S\ref{sec_1dfitting} are subject to bias if the tangential shear profile around one halo is significantly influenced by the presence of the other. Thus, for more accurate mass measurement, we must fit two halos simultaneously. 

As in \S\ref{sec_1dfitting}, we assume that each cluster's center coincides with the location of the BCG and fix the centroids of both halos in our two-dimensional fitting. 
Because each subcluster's mass is not large enough to produce strong signals that can constrain six free parameters (two centroid coordinates and two concentrations), fixing the centroid is required to stabilize the fitting. Since 
the WL mass peaks, the $Chandra$ X-ray peaks, and the BCG positions agree excellently, it is unlikely that this centroid choice leads to any significant mass estimate bias.
We model both halos with NFW profiles using the mass-concentration of \cite{Dutton2014} and determine the expected shear at every source galaxy position based on the combined contribution from the halos. Our log-likelihood is given as
\begin{equation}
L = \sum_{i} \sum_{s=1,2} \frac{ [ g^m_s (M_{A115N},M_{A115S},x_i,y_i) - g^o_s(x_i,y_i)]^2 } {\sigma_{SN}^2 + \sigma_e^2},
\end{equation}
\noindent
where $g^m_s$ ($g^o_s$) is the $s^{th}$ component of the predicted (observed) reduced shear at the $i^{th}$ galaxy position $(x_i,y_i)$ as a function of the two clusters' masses $M_{A115N}$ and $M_{A115S}$.  
The ellipticity dispersion (shape noise) is $\sigma_{SN}=0.25$ whereas $\sigma_e$ is the ellipticity measurement noise of each object. Note that the evaluation of this likelihood function does not require source galaxy binning.

\begin{figure}[h!]
        \centering
         \includegraphics[width=0.46\textwidth]{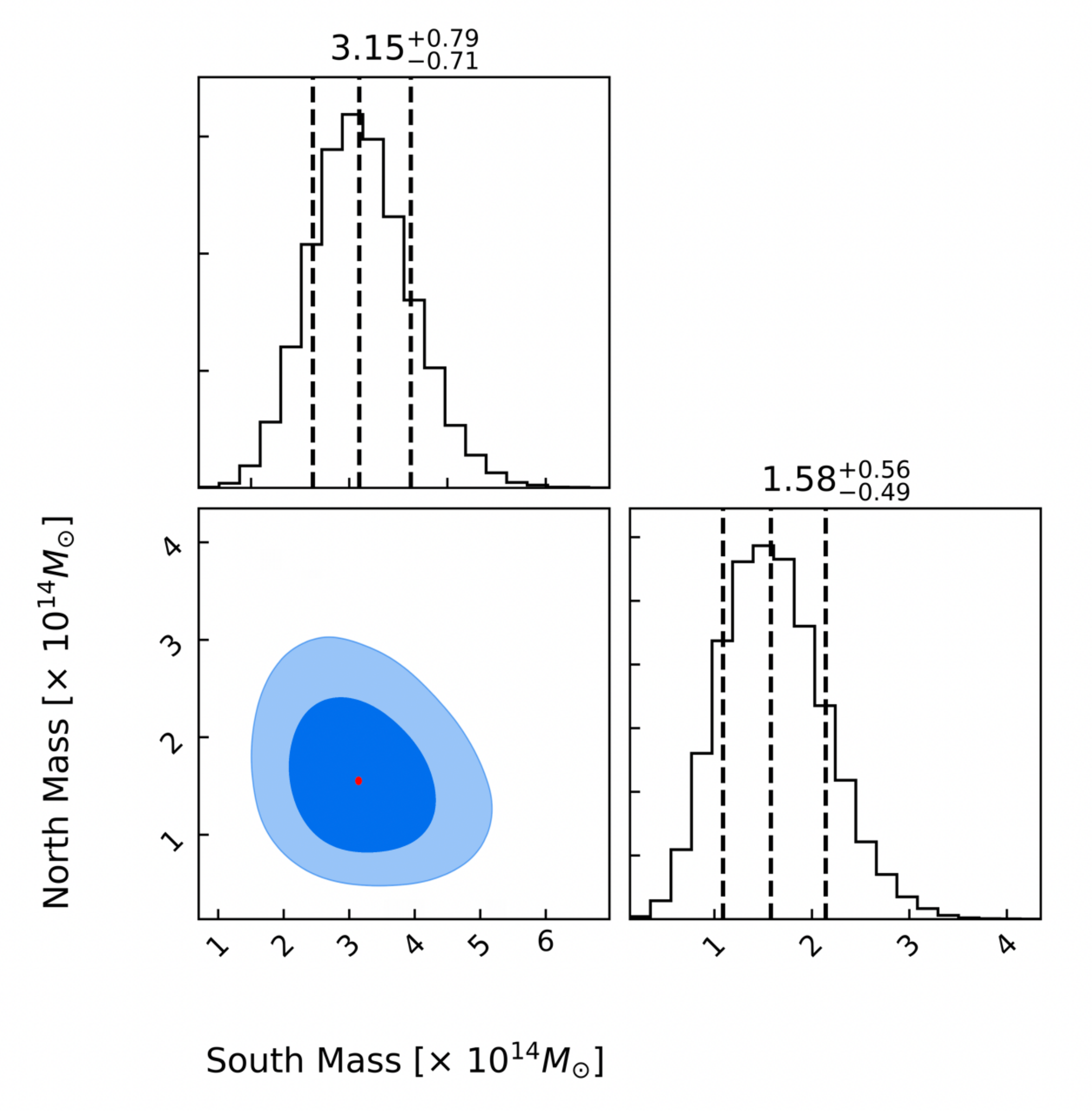}
        \caption{Mass determination of A115N and A115S from our simultaneous two-dimensional fitting with two NFW halos.
         We used the MCMC sampling method to explore the parameter space. The result shown here is derived from one million chains.
         Since we assume the mass-concentration relation of \cite{Dutton2014}, the total number of free parameters is two.
         The dashed lines indicate the 1$\sigma$ uncertainties while the inner and outer contours show the 1$\sigma$ and 2$\sigma$ regions, respectively. The degeneracy between the two subcluster masses is weak (the Pearson correlation coefficient is $\rho=-0.256$).
        } 
        \label{fig:mcmc_2par}
\end{figure}

We used the Markov-Chain-Monte-Carlo (MCMC) method to sample this likelihood.
We display the resulting parameter contours in Figure \ref{fig:mcmc_2par} and list the best-fit parameters in Table~\ref{tab:mcmc_2par}. One may expect a degeneracy between the two parameters to exist to some extent because the two masses can trade with each other without significantly affecting the global goodness-of-the-fit. However, we find that the degeneracy is weak, which is attributed to the large distance ($\mytilde900$~kpc) between the two subclusters.
The masses for A115N and A115S are $M_{200c}=1.58_{-0.49}^{+0.56}\times10^{14}M_{\sun}$ and $3.15_{-0.71}^{+0.79}\times10^{14}M_{\sun}$, respectively.
These masses are consistent with our one-dimensional fitting results, although the decrease in the central value is in line with our expectation.

The total mass of A115 is not a simple sum of the two masses of A115N and A115S if we maintain a consistent scheme in defining halo masses (i.e., a mass contained within a spherical volume as defined in \S\ref{sec_introduction}). In order to derive the total $M_{200c}$  mass, we need to determine $R_{200c}$ for the total system, which requires the two following assumptions. First, we assume that the two subclusters are merging on the plane of the sky. This allows us to adopt the projected distance as the physical separation between A115N and A115S. Since our analysis (\S\ref{sec_discussion}) favors the scenario wherein the merger is happening nearly in the plane of the sky, we believe that this assumption does not greatly depart from the truth. Second, we assume that the system's global center is located at the geometric mean of the two subclusters. One may argue that a better choice would be the barycenter. However, our analysis shows that this change causes a less than 10\% shift in the total mass. We populate a three-dimensional grid with the sum of two densities based on the NFW parameters of both clusters. The $R_{200c}$ value is determined by locating the radius of the spherical volume, inside which the mean density becomes 200 times the critical density of the universe at the cluster redshift. The total mass obtained in this way is  $M_{200c}=6.41_{-1.04}^{+1.08}\times10^{14}M_{\sun}$ at $R_{200c}=1.67_{-0.09}^{+0.10}$~Mpc.
Comparison of these WL masses with our X-ray and spectroscopic results and the values in the literature are discussed in \S\ref{sec_comparison}.

\begin{table}[]
\centering
\caption{2D Two Halo NFW Fitting Result}
\label{tab:mcmc_2par}
\begin{tabular}{ccc}
\hline \hline
  & \begin{tabular}[c]{@{}c@{}}$R_{200c}$\\ (Mpc)\end{tabular} & \begin{tabular}[c]{@{}c@{}}$M_{200c}$\\ $(\times10^{14} M_{\sun})$\end{tabular} \\ \hline
Global & $1.67_{-0.09}^{+0.10}$ & $6.41_{-1.04}^{+1.08}$ \\
North & $1.03^{+0.13}_{-0.11}$ & $1.58^{+0.56}_{-0.49}$ \\
South & $1.31^{+0.11}_{-0.10}$ & $3.15^{+0.79}_{-0.71}$ \\ \hline \hline
\end{tabular}
\end{table}

\subsection{X-ray Mass Estimation}

Our first step toward X-ray-based mass estimation is the measurement of the ICM temperature. We used the X-ray spectra within a 1-5 keV energy band and the MEKAL plasma model \citep{Kaastra1993,Liedahl1995}. Exclusion of the energy band less than 1~keV is our conservative measure to minimize the impact from the well-known low-energy calibration issue of the ACIS detector \citep[e.g.,][]{chartas2002}.
The Galactic hydrogen density, metallicity abundance, and redshift of the cluster were fixed to $N_{H}=5.2 \times 10^{20} \mbox{cm}^{-2}$ \citep{Stark1992}, $Z_{\sun}=0.3$, and $z=0.192$, respectively. Because each X-ray peak possesses a disturbed morphology and a position-dependent temperature variation, it is difficult to determine a single temperature representative of the cluster mass. We took care to avoid using too small an aperture because each peak has a cool core and using too large an aperture because the intermediate region between the two X-ray peaks has a very high temperature, which is attributed to the on-going merger activity \citep{Hallman2018}. We used the temperature map of \cite{Hallman2018} as a guide. The resulting inner and outer radii of our  annuli's are 94 kpc and 283 kpc (94 kpc and 267 kpc), respectively for A115N (A115S).

\begin{figure*}
        \centering
        \includegraphics[width=0.9\textwidth]{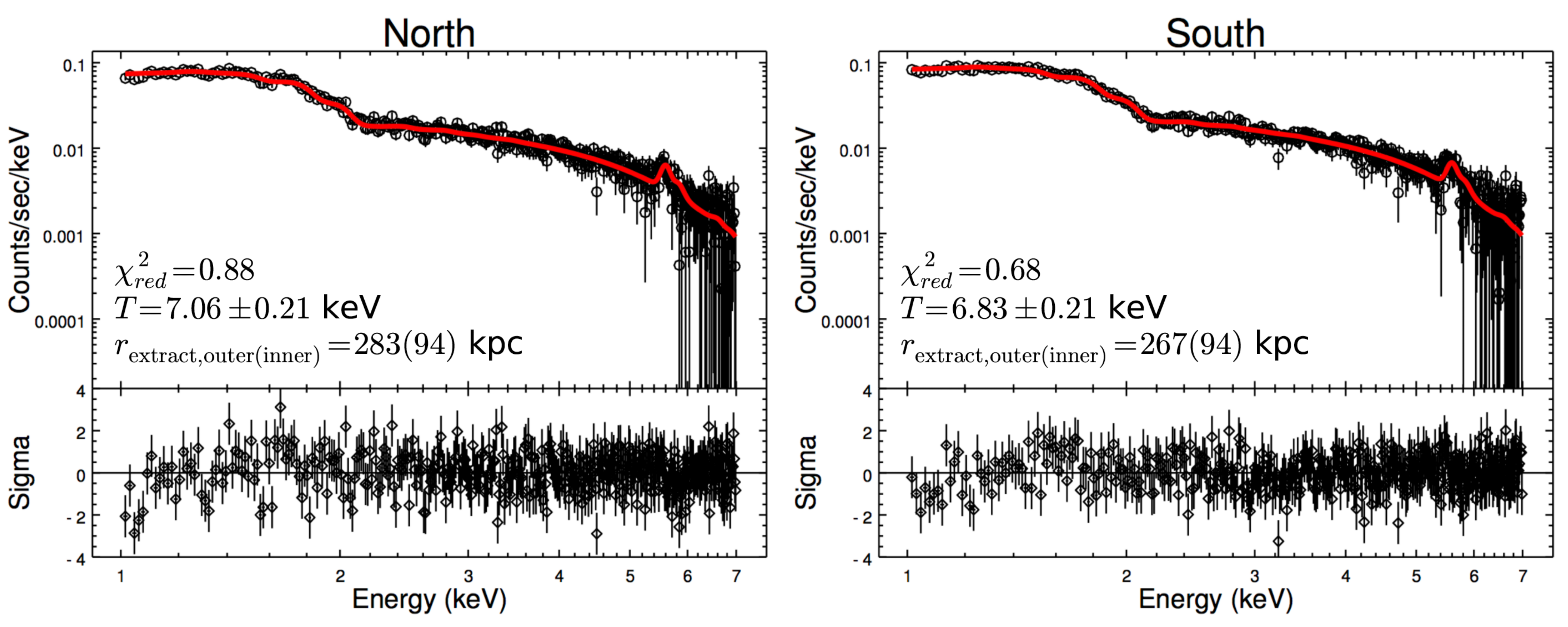}
        \caption{Core-excised {\it Chandra} X-ray spectra of A115N and A115S. The upper boxes show the spectra whereas the lower boxes display the residuals. The red solid lines represent the best-fit results based on the MEKAL model.} 
        \label{fig:xray_spectra}
\end{figure*}

Using the above setup, we measured $T_X=7.06 \pm 0.21$~keV ($\chi^2_{red}=0.88$) and $6.83 \pm 0.21$~keV ($\chi^2_{red}=0.68$) for A115N and A115S, respectively. We display the X-ray spectra and fitting results in Figure~\ref{fig:xray_spectra}. 
If we do not excise the cores, the temperatures become $T_X=5.08 \pm 0.08$ keV and $T_X=6.86 \pm 0.19$ keV for A115N and A115S, respectively. The decrease in A115N is significant and shows that the core temperature of A115N is indeed low.
As shown by previous studies, we also confirm that using a smaller circular aperture leads to lower temperatures for both X-ray peaks. For example, choosing an $r=47$~kpc aperture gives $T_X=3.19 \pm 0.06$~keV for A115N and $5.12 \pm 0.51$~keV for A115S. These measurements are consistent with the measurements in \cite{Gutierrez2005}.

One popular method for X-ray-based mass estimation is to determine the mass using both X-ray and surface brightness measurements with the assumption that the halo follows a certain analytic profile such as NFW. We do not employ this method here, however, because the disturbed morphology prevents us from obtaining a reliable surface brightness profile. Instead, we estimate the cluster mass from a mass-temperature ($M-T$) relation based on the temperature measurements extracted from the aforementioned annuli.

Using the scaling relations of  \cite{Mantz2016} gives $M_{500c}=6.29^{+1.39}_{-1.01}\times10^{14}M_{\sun}$ and $5.96^{+1.30}_{-0.95}\times10^{14}M_{\sun}$ for A115N and A115S, respectively.
For comparison with weak-lensing masses, we converted these $M_{500c}$ masses to $M_{200c}$ masses by extrapolation. Using the mass-concentration relation of \cite{Dutton2014}, we obtained $M_{200c}=9.00^{+2.03}_{-1.48}\times10^{14}M_{\sun}$ ($R_{200c}=1.87^{+0.13}_{-0.11}$ Mpc) for A115N and $8.52^{+1.90}_{-1.38}\times10^{14}M_{\sun}$ ($R_{200c}=1.84^{+0.13}_{-0.10}$ Mpc) for A115S. As mentioned in \S\ref{sec_wl_mass_estimation}, the total mass of A115 is not a simple sum of the two masses. Using the method described in \S\ref{sec_wl_mass_estimation}, we estimated the total X-ray mass to be  $M_{200c}=20.48^{+3.49}_{-2.71}\times10^{14}M_{\sun}$ or $M_{500c}=14.06^{+2.34}_{-1.82}\times10^{14}M_{\sun}$.

\begin{table*}[]
\centering
\caption{X-ray Mass}
\label{tab:x_ray_mass}
\begin{tabular}{ccccc}
\hline \hline
X-ray  & \begin{tabular}[c]{@{}c@{}}$R_{500c}$\\ $(\mbox{Mpc})$\end{tabular} & \begin{tabular}[c]{@{}c@{}}$R_{200c}$\\ $( \mbox{Mpc})$\end{tabular} & \begin{tabular}[c]{@{}c@{}}$M_{500c}$\\ $(\times 10^{14} M_{\sun})$\end{tabular} & \begin{tabular}[c]{@{}c@{}}$M_{200c}$\\ $(\times 10^{14} M_{\sun})$\end{tabular} \\ \hline
Global & $1.60^{+0.08}_{-0.07}$    & $2.46^{+0.13}_{-0.12}$     & $14.06^{+2.34}_{-1.82}$   & $20.48^{+3.49}_{-2.71}$   \\
North  & $1.22^{+0.08}_{-0.07}$    & $1.87^{+0.13}_{-0.11}$    & $6.29^{+1.39}_{-1.01}$   & $9.00^{+2.03}_{-1.48}$   \\
South  & $1.20^{+0.08}_{-0.07}$    & $1.84^{+0.13}_{-0.10}$    & $5.96^{+1.30}_{-0.95}$   & $8.52^{+1.90}_{-1.38}$ \\ \hline \hline
\end{tabular}
\end{table*}

%\subsection{Dynamical Mass Estimation}
\subsection{Dynamical Mass Estimation} 
\label{sec:dynamical_mass_estimation}
We compiled our spectroscopic redshift galaxy catalog of the A115 field by combining the \cite{Golovich2017} and \cite{Rines2018} data. The \cite{Golovich2017} catalog contains 198 spectroscopic members from our own DEIMOS survey and NASA/IPAC Extragalactic Database (NED)\footnote{https://ned.ipac.caltech.edu}. The NED catalog has contributions from \cite{Beers1990}, \cite{Zabludoff1990}, \cite{Barrena2007}, \cite{Skrutskie2006}, and \cite{Alam2015}. From their HeCS-red survey, \cite{Rines2018} provided 512 objects in the A115 field, of which 95 are A115 members. Out of these 95 objects, 27 are redundant with those in the \cite{Golovich2017} catalog. We verified that the spectroscopic redshifts of these 27 common objects agree excellently to the fourth decimal point. The total number of A115 cluster members in our combined catalog is 266. 

\begin{figure}[h!]
        \centering
        \includegraphics[width=0.46\textwidth,trim=3cm 0cm 3cm 0cm]{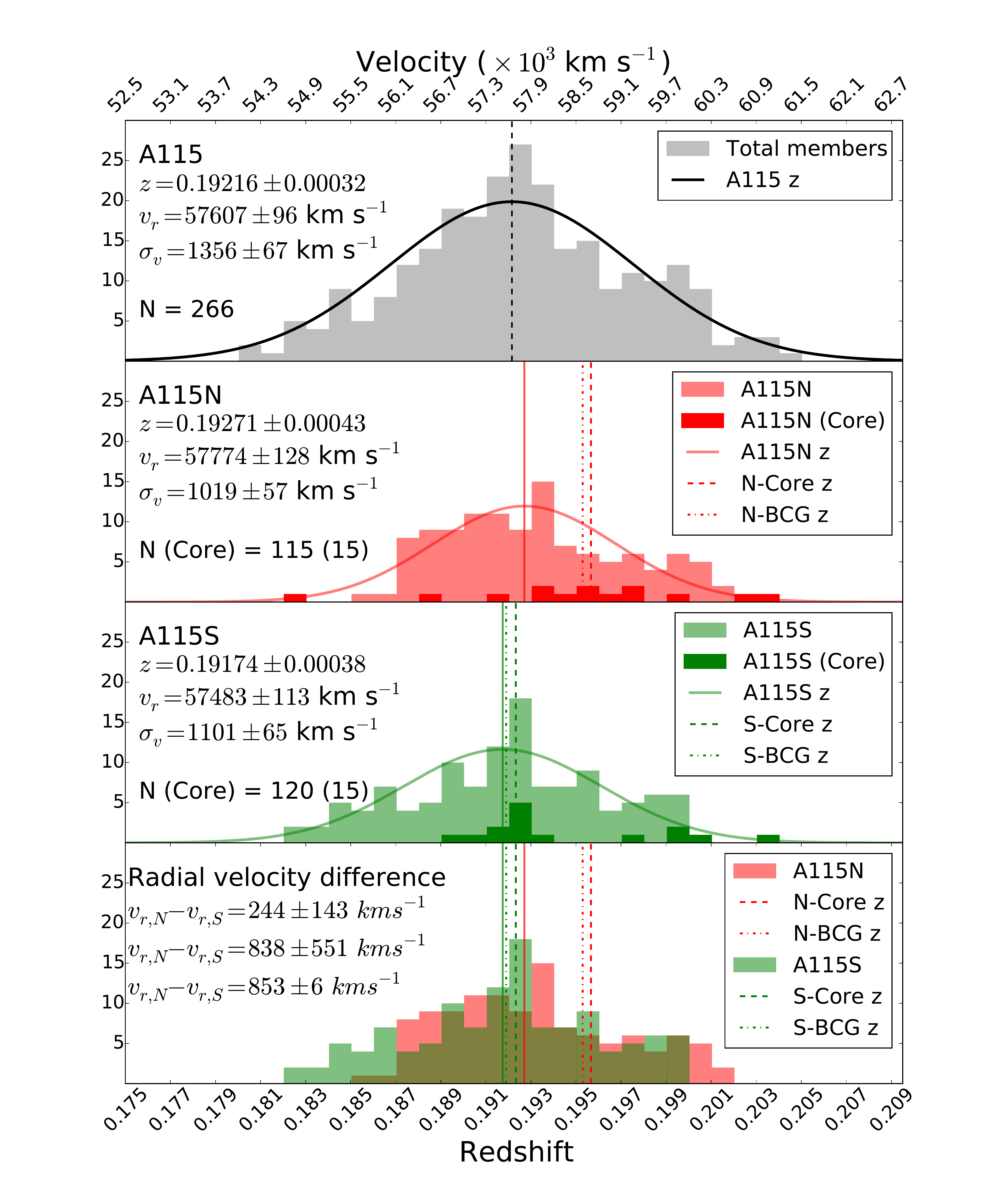}
        \caption{Redshift distribution of 266 cluster member galaxies and velocity dispersion estimation. The top panel shows the global redshift distribution. The second and third panels represent the redshift distribution of the northern and southern subclusters, respectively. The bottom panel shows the radial velocity differences of the subclusters, core regions, and BCGs. The membership was determined by the Gaussian Mixture Model (GMM). The members within the core region of 0.25 $h^{-1}$ Mpc radius are represented by dark shades. We performed $\sigma$-clipping on both subcluster members to remove the outliers. The means and standard deviations of overlaid Gaussians are from the biweight statistics \citep{Beers1990}. The radial velocity $v_r$ is measured from the classical Doppler effect relation $v_r=cz$, where $c$ is the speed of light.
        The velocity dispersion is measured in the rest frame of the cluster. The solid, dashed, and dotted lines on each panel are the mean redshift of the cluster, core region, and the redshift of BCG, respectively.   } 
        \label{fig:vel_dist}
\end{figure}

We applied the bi-weight estimator \citep{Beers1990} and determined the redshift and LOS velocity dispersion of A115 to be $z=0.19216\pm0.00032$ and $\sigma_{v}=1356 \pm 67~\mbox{km}~\mbox{s}^{-1}$, respectively; we use bootstrapping to evaluate the uncertainties.
Both values are consistent with the \cite{Barrena2007} measurements ($z=0.1929\pm0.0005$ and $\sigma_{v}=1362_{-108}^{+126}~\mbox{km}~\mbox{s}^{-1}$) and also with the \cite{Golovich2018} results ($z=0.19285\pm0.00040$ and $\sigma_{v}=1439\pm79~\mbox{km}~\mbox{s}^{-1}$). The top panel of Figure \ref{fig:vel_dist} shows the redshift distribution of the 266 members of A115. We agree with \cite{Golovich2017} that the overall redshift distribution of the A115 galaxies is well-described with a single Gaussian profile.

Assigning a galaxy to one of the two subclusters is non-trivial because their virial radii overlap. We used a Gaussian Mixture Model (GMM) analysis\footnote{We used the {\tt scikit-learn} implementation available at https://scikit-learn.org/stable/modules/mixture.html.} to determine the membership between A115N and A115S. The analysis assigned 134 and 132 galaxies to A115N and A115S, respectively. After $\sigma$-clipping,
the number of members reduced to 115 (120) for A115N (A115S).
The second and third panels (light shade) of Figure \ref{fig:vel_dist} display the redshift distributions of A115N and A115S, respectively.
The LOS difference in velocity between the two subsystems is $244\pm144~\mbox{km}~\mbox{s}^{-1}$ (see the bottom panel of Figure \ref{fig:vel_dist}). 
The individual velocity dispersions of A115N and A115S are $\sigma_{v}=1019\pm57~\mbox{km}~\mbox{s}^{-1}$ and $\sigma_{v}=1101\pm64~\mbox{km}~\mbox{s}^{-1}$, respectively.

We converted the above velocity dispersions to dynamical masses using the $M-\sigma_v$ scaling relation of \cite{Saro2013}. The dynamical mass of the entire system was estimated to be $M_{200c}=37.4_{-5.2}^{+5.7}\times 10^{14}M_{\sun}$ while we obtained $M_{200c}=16.3_{-2.5}^{+2.8}\times 10^{14}M_{\sun}$ and $M_{200c}=20.4_{-3.3}^{+3.7}\times 10^{14}M_{\sun}$ for A115N and A115S, respectively.

\cite{Barrena2007} quoted a very large ($\mytilde1600~\mbox{km}~\mbox{s}^{-1}$) velocity difference between A115N and A115S from their analysis of 88 cluster members. This claim is based on measurement of only the members within $\mytilde0.25$ Mpc of the BCG. However, this measurement lacks statistical significance because only 6 and 7 members for A115N and A115S, respectively, were found within this radius. When we repeated the analysis using our catalog, we found 15 members for each subcluster within the same radius. The redshift distribution of these galaxies are shown as the dark shaded histograms in the second and third panels of Figure \ref{fig:vel_dist}. The LOS velocity difference measured in this way is $838\pm551~\mbox{km}~\mbox{s}^{-1}$. The central value is higher than the case where we use the GMM method to determine the subcluster membership ($244\pm144~\mbox{km}~\mbox{s}^{-1}$). However, the two measurements are different only by $\mytilde1\sigma$ because of the large uncertainty attached to the measurement from the members in the subcluster core. Nevertheless, it is interesting to note that the LOS velocity difference between the two BCGs is $\mytilde853~\mbox{km}~\mbox{s}^{-1}$, which is close to the central value of the measurement $838\pm551~\mbox{km}~\mbox{s}^{-1}$ based on the members in the core ($r<0.25$ Mpc).
The change in the LOS velocity happens mostly because the galaxies located in the A115N center on average have higher redshifts than the rest (see the solid versus dashed lines in the second panel of Figure~\ref{fig:vel_dist}). We do not observe this trend for A115S (the third panel of Figure~\ref{fig:vel_dist}).
This radial dependence is also mentioned by \cite{Barrena2007} in Figure 13 and 14 of their paper.   
We defer our interpretation of the above results to \S\ref{sec_merging_scenario}.

\subsection{$M/L$ Ratio Estimation}

Mass-to-light ratios ($M/L$) of galaxy clusters have been used to estimate the matter density of the universe under the assumption that clusters are representative of our universe \citep[e.g.,][]{carlberg1997}. Also, the evolution of cluster $M/L$ values with redshift provide useful constraints on the stellar mass assembly history. Here we present our estimation of the $M/L$ value of A115. One of the motivations of this investigation is to examine the consistency of the resulting $M/L$ values with results for other clusters. Although the $M/L$ dispersion among clusters is quite large in the literature (for example, the $M/L$ value spans the range $50\sim1000$ for $10^{14}-10^{15} M_{\sun}$ clusters according to Girardi et al. 2002), the order-of-magnitude difference between our WL and other mass estimates makes this comparison still statistically interesting. To measure the $M/L$ value, we evaluated the A115 mass and luminosity within a cylindrical volume rather than a spherical volume. 
We used the best-fit NFW parameters presented in our two-halo simultaneous fitting  (\S\ref{sec:two_halo_fitting}) to estimate the projected mass density as a function of radius. The projected mean surface mass density for each subcluster was computed using Equation 13 from \cite{Wright2000}. A two-dimensional mass density map was obtained by adding the contributions from A115N and A115S.

We constructed our A115 member catalog by combining our spectroscopic members and photometrically selected member candidates based on the color-magnitude relation. We characterized the red-sequence locus by performing a linear fit to the spec-$z$ members and selected the candidate galaxies whose $V-i^{\prime}$ colors are within 0.05 magnitude from the fitted line and $V$-band magnitudes are brighter than $V=22$  (see photometric candidate in Figure~\ref{fig:CMD}). Our final member catalog contains 377 objects. 
We estimated $B$-band luminosity $L_{B\sun}$
from our $V$ and $i^{\prime}$ magnitudes using the photometric transformation obtained by performing synthetic photometry \citep{Sirianni2005} with a spectral energy distribution (SED) template of elliptical galaxies. 

\begin{figure}
        \centering
        \includegraphics[width=0.46\textwidth]{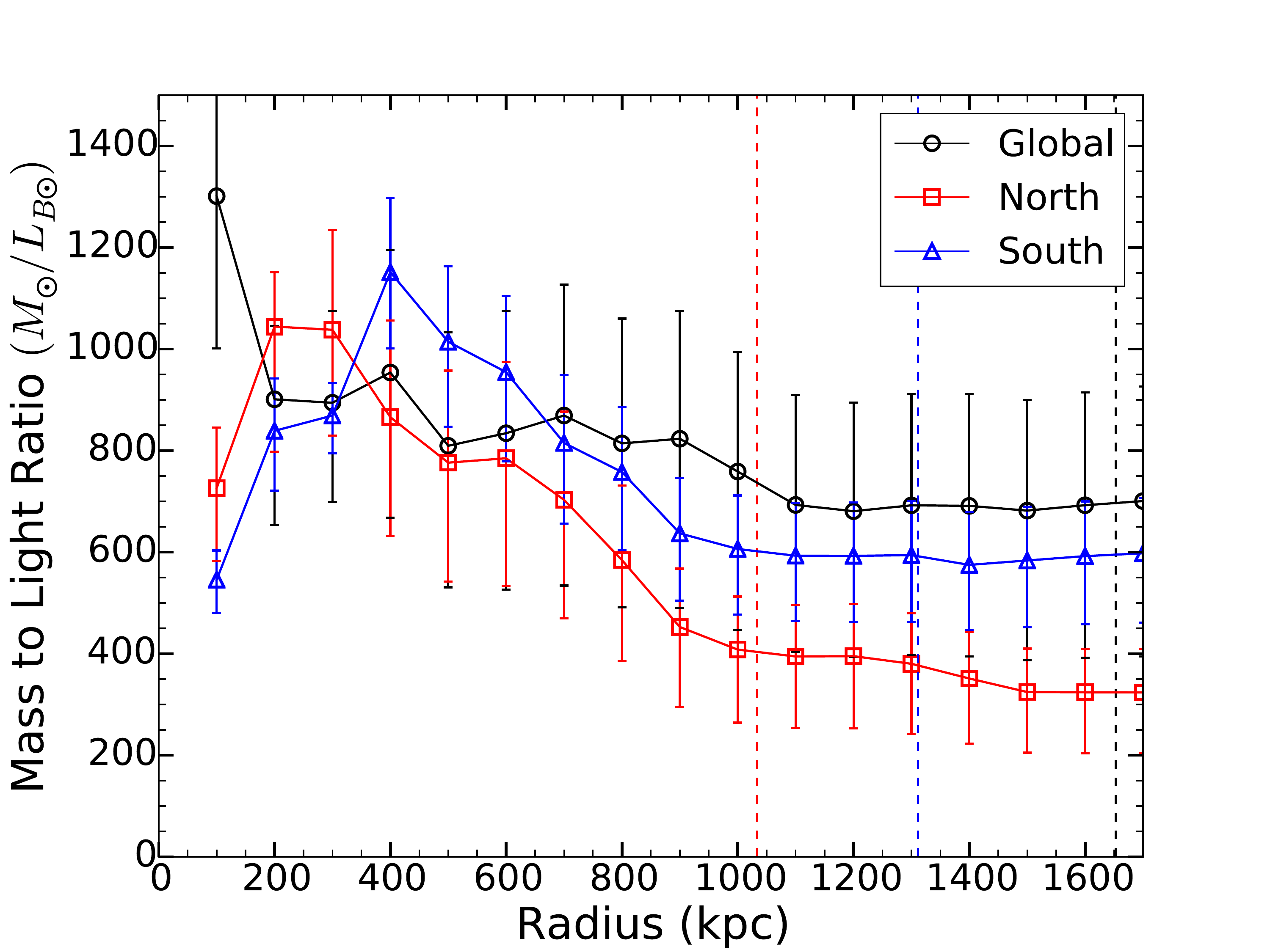}
        \caption{Cumulative $M/L$ profile of A115. Black, red, and blue open markers represent the $M/L$ ratios of the global, northern, and southern clusters, respectively. The vertical dashed lines indicate the virial radius of each cluster. In this plot, we use projected masses derived from our NFW model and galaxy luminosity from a photometrically selected red sequence.}
        \label{fig:MLratio}
\end{figure}

Figure \ref{fig:MLratio} shows the cumulative $M/L$ profile for our three chosen centers (two BCGs and one global). When the centers are placed at the BCGs, the $M/L$ value is low at small radii because of the BCG's contribution to the luminosity. The $M/L$ value is estimated high near the global center because no bright galaxies are present in this region. 
We find that the $M/L$ ratio of A115N and A115S are $\mytilde400$ and $\mytilde650$, respectively within their virial radii ($\mytilde1$~Mpc for A115N and $\mytilde1.3$~Mpc for A115S). These $M/L$ values are higher than the mean value of the $\Lambda$CDM prediction, but can be accommodated
within the distribution of the sample of 89 clusters studied in \cite{Girardi2002}. This comparison shows that our WL masses, although substantially lower than the X-ray or dynamical estimates, give the most physical $M/L$ values for A115. If dynamical masses are used instead, the implied $M/L$ value would increase by an order of magnitude, which is difficult to accommodate within the current $\Lambda$CDM paradigm. In general, dynamics of galaxies are known to be biased in a merger \citep{Pinkney1996,Takizawa2010}.

\section{Discussion}
\label{sec_discussion}

\subsection{Comparison with Previous Mass Estimates}
\label{sec_comparison}
A115 is one of the most studied galaxy clusters. Here, we compare our WL mass estimates with those from the literature.

\begin{figure}[h!]
        \centering
        \includegraphics[width=0.46\textwidth]{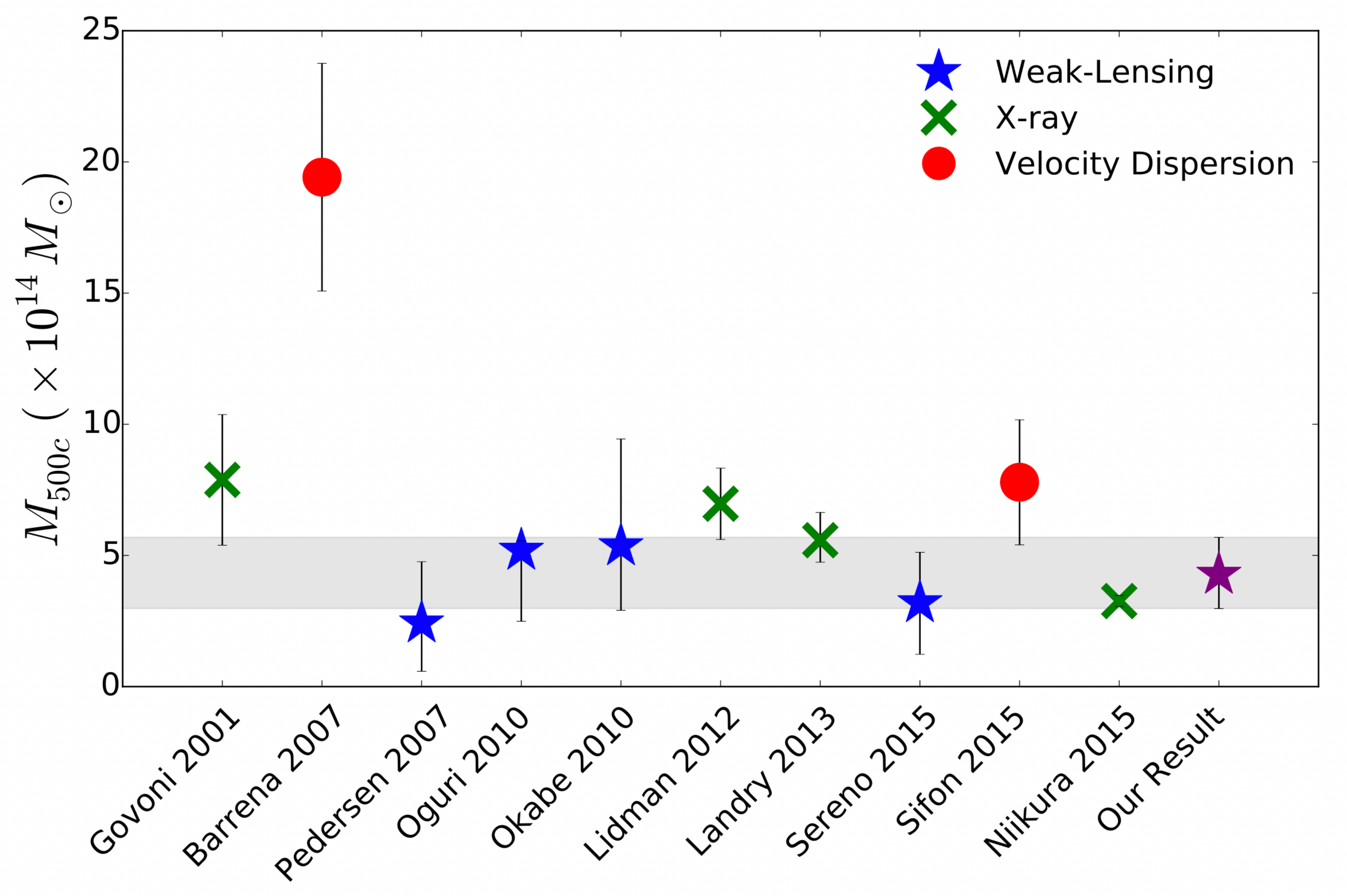}
        \caption{Global mass estimations of A115 from previous research. Global $M_{500c}$ of the cluster is compared on the plot with $1\sigma$ uncertainty error bars. In the case that the previous research only measured $M_{200c}$, we converted $M_{200c}$ to $M_{500c}$ assuming the cluster follows the NFW halo model. The gray shaded region is the error region of our mass estimation. The results are sorted in chronological order. Our mass estimation is $1\sigma$-consistent with other weak-lensing masses. Dynamical and X-ray masses tend to be higher than the weak-lensing masses, which we attribute to their assumption of the cluster being in hydrostatic equilibrium.
        }
        \label{fig:mass_literature}

\end{figure}

{\bf Global Mass.} Figure~\ref{fig:mass_literature} shows the global $M_{500c}$ estimates from various studies. Note that most past studies did not report separate masses for A115N and A115S. For studies that only quote $M_{200c}$ values, we converted them to $M_{500c}$ values using an NFW profile and the mass-concentration relation of \cite{Dutton2014}. This conversion was also applied to our WL results.
 
The most significant outlier in Figure~\ref{fig:mass_literature} is the dynamical mass estimate from \cite{Barrena2007}. Because our velocity dispersions from improved statistics yield a similarly high mass, we attribute the large difference not to any errors in measurement, but to a significant departure of A115 from dynamical equilibrium due to the merger. 
In general, velocity dispersion is believed to be boosted in epochs close to pericentric passages \citep[e.g.,][]{Pinkney1996,Monteiro-Oliveira2017}.
However, it remains to be investigated by future numerical simulations whether or not the merger alone can inflate the velocity dispersion measurement to this extent. 
The dynamical mass estimate from \cite{Sifon2015} is substantially lower than the \cite{Barrena2007} result. This is because \cite{Sifon2015} treated A115 as a single halo whereas \cite{Barrena2007} took into account the multiplicity.

The X-ray and WL mass estimates presented in Figure~\ref{fig:mass_literature} seem to be consistent with our weak lensing result. However, the caveat is that these values are obtained under the single-halo assumption.

{\bf Substructure Mass.} 
\cite{Hoekstra2012} presented WL masses for A115N and A115S separately using Canada-France-Hawaii Telescope (CFHT) imaging data.
They quoted $M_{500c}=3.9_{-1.5}^{+1.4}\times 10^{14} M_{\sun}$ and $5.4_{-1.2}^{+1.3} \times 10^{14} M_{\sun}$ for A115N and A115S, respectively. These masses were derived by de-projecting their aperture masses. When they directly fit an NFW profile, they obtain $M_{500c}=3.2_{-1.0}^{+1.0}\times 10^{14} M_{\sun}$ ($3.8_{-1.1}^{+1.2} \times 10^{14} M_{\sun}$ ) for A115N (A115S).
The de-projected values are higher than our results by a factor of 2-3; when converted to $M_{500c}$, our Subaru-base WL masses become $M_{500c}=1.14^{+0.40}_{-0.35}\times10^{14}M_{\sun}$ and $2.25^{+0.55}_{-0.50}\times10^{14}M_{\sun}$ for A115N and A115S, respectively.
When the two masses (A115N and A115S) from \cite{Hoekstra2012} are combined, the resulting global mass of A115 would be also 2-3 times higher than our WL result.
In order to investigate the source of the discrepancy with the \cite{Hoekstra2012} results, we analyzed their CFHT data with our WL pipeline. The difference in depth and seeing results in a slight ($\mytilde30$\%) reduction in source density compared to the Subaru analysis ($\mytilde19~\mbox{arcmin}^{-2}$ vs $\mytilde24~\mbox{arcmin}^{-2}$). Nevertheless, we find that our masses derived from the CFHT data are in agreement with our Subaru-based values within $\mytilde2$\%. This excellent agreement supports the repeatability of our WL mass measurement regardless of the instrument choice.
Hence, we suspect that the discrepancy between \cite{Hoekstra2012} and ours may be attributed to the difference in the WL pipeline and mass estimation method.

{\bf Mass distribution.} 
Among the few WL studies in the literature, only \cite{Okabe2010} presented a two-dimensional mass distribution for A115, which shows two mass peaks similar to ours. 
However, both of their mass clumps are offset toward the northeast with respect to their nearest BCGs. As mentioned in \S\ref{sec_WL_analysis}, our mass peaks coincide with the corresponding BCGs. \cite{Okabe2010} performed their WL analysis using the $i^{\prime}$-band image, which was significantly deeper than the $V$-band image at the time of the analysis. Because our WL shape is derived from the $V$-band data, we think that the difference may be due to different systematics. To address the issue, we repeated the measurement with the $i^{\prime}$ imaging data. We find that the position-dependent PSF ellipticity pattern of the $i^{\prime}$ image is much more complex than the pattern in the $V$ image and our PCA-based PSF model could not reproduce the observed PSF pattern with the same fidelity (Figure~\ref{fig:PSF_com}), as mentioned in \S\ref{sec_psfmodeling}. Interestingly, the resulting mass reconstruction from this $i^{\prime}$-band analysis resembles the one in \cite{Okabe2010}, possessing similar offsets. Therefore, it is possible that the mass-galaxy offsets in \cite{Okabe2010} may be due to large residual PSF systematics in the $i^{\prime}$-band imaging data. However, we can only be speculative regarding this issue because we do not have access to their WL catalog.

\begin{table}[]
\centering
\caption{Mass Comparison}
\label{tab:mass_com}
\begin{tabular}{cccc}
\hline \hline
$M_{200c}$\\ $(\times 10^{14} M_{\sun})$ & Global                  & North                     & South                   \\ \hline
Weak Lensing                                                                    & $6.41^{+1.08}_{-1.04}$  & $1.58^{+0.56}_{-0.49}$    & $3.15^{+0.79}_{-0.71}$  \\
X-ray                                                                           & $20.48^{+3.49}_{-2.71}$ & $9.00^{+2.03}_{-1.48}$ & $8.52^{+1.90}_{-1.38}$ \\
Velocity Dispersion                                                             & $37.4^{+5.7}_{-5.2}$ & $16.3^{+2.8}_{-2.5}$   & $20.4^{+3.7}_{-3.3}$\\
\hline \hline 
\end{tabular}
\end{table}

\subsection{Significance of Weak-lensing Mass Centroid} \label{sec_offset}

\begin{figure*}
        \centering
        \includegraphics[width=0.9\textwidth]{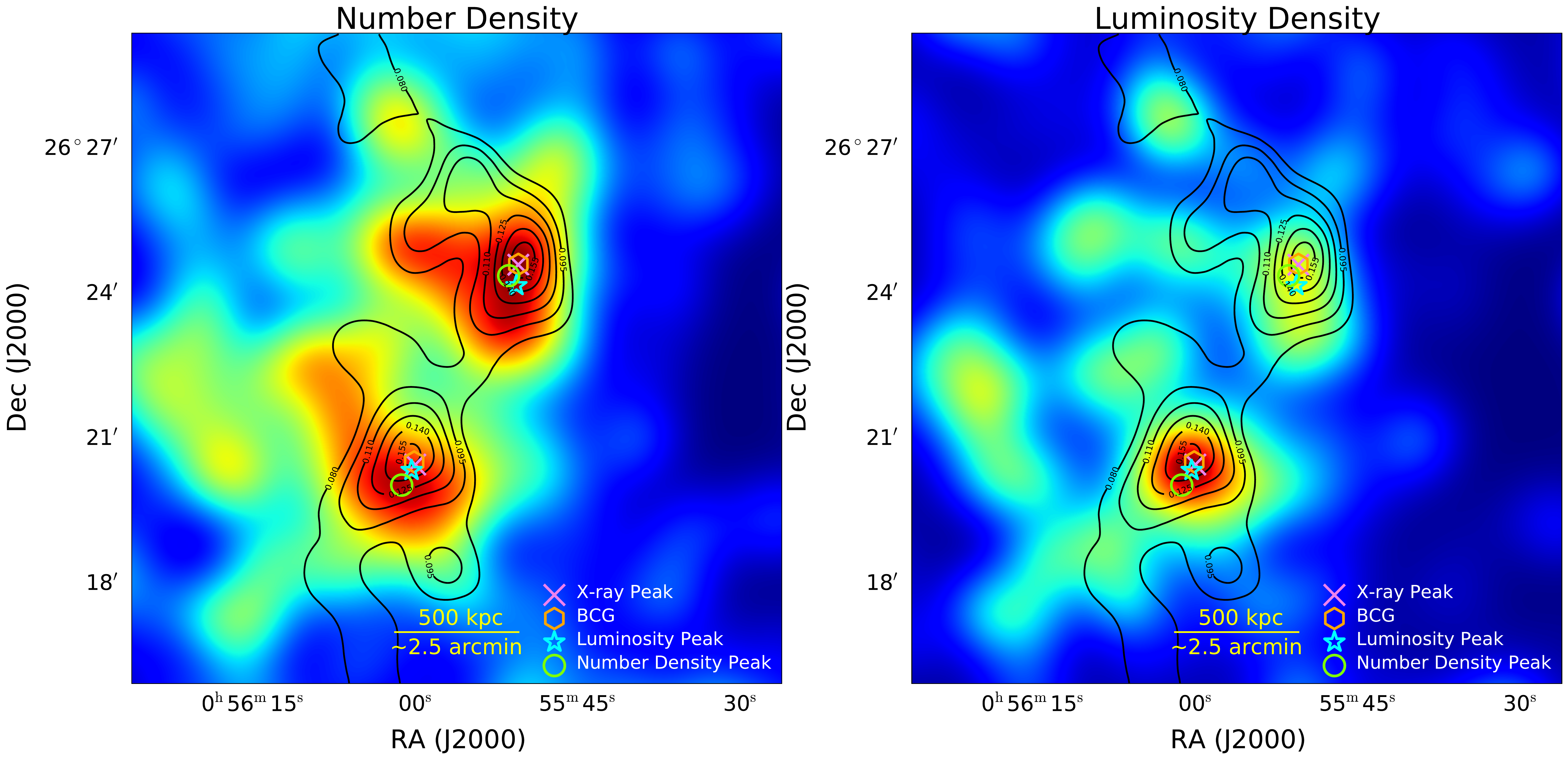}
        \caption{Convergence overlaid on the number and luminosity density maps of the cluster members. A total of 377 cluster members (266 spectroscopic members and 111 photometric members) are used to create these number and luminosity maps. The displayed results are obtained after smoothing with a FWHM $=188\arcsec$ Gaussian kernel. Our bootstrapping analysis (see text) shows that the five centroids (mass, X-ray, galaxy number, galaxy luminosity, and BCG location) are statistically consistent.}
        \label{fig:lum_num_den}

\end{figure*}

As shown in Figure~\ref{fig:convergence_map}, our mass centroids agree nicely with the BCG positions. If the BCG represents the true center of each halo, one can interpret the agreement as evidence for dark matter with negligible self-interacting cross-section. However, it is still unclear in general whether or not a BCG can serve as the proxy for a halo center.
Alternatively, one can use smoothed galaxy distributions to define halo centers. In Figure \ref{fig:lum_num_den}, we display mass contours over galaxy number and luminosity density maps. 
Interestingly, the centroids of the smoothed galaxy distributions possess offsets with respect to the BCGs. For A115S, both number and luminosity density centroids are displaced south by $\mytilde30\arcsec$. Similar offsets are found for A115N except that the number density peak is at a greater distance from the BCG than the luminosity peak. Here we present our investigation of the statistical significance of the mass centroid with respect to various definitions of subcluster centers.

\begin{figure}
        \centering
        \includegraphics[width=0.46\textwidth]{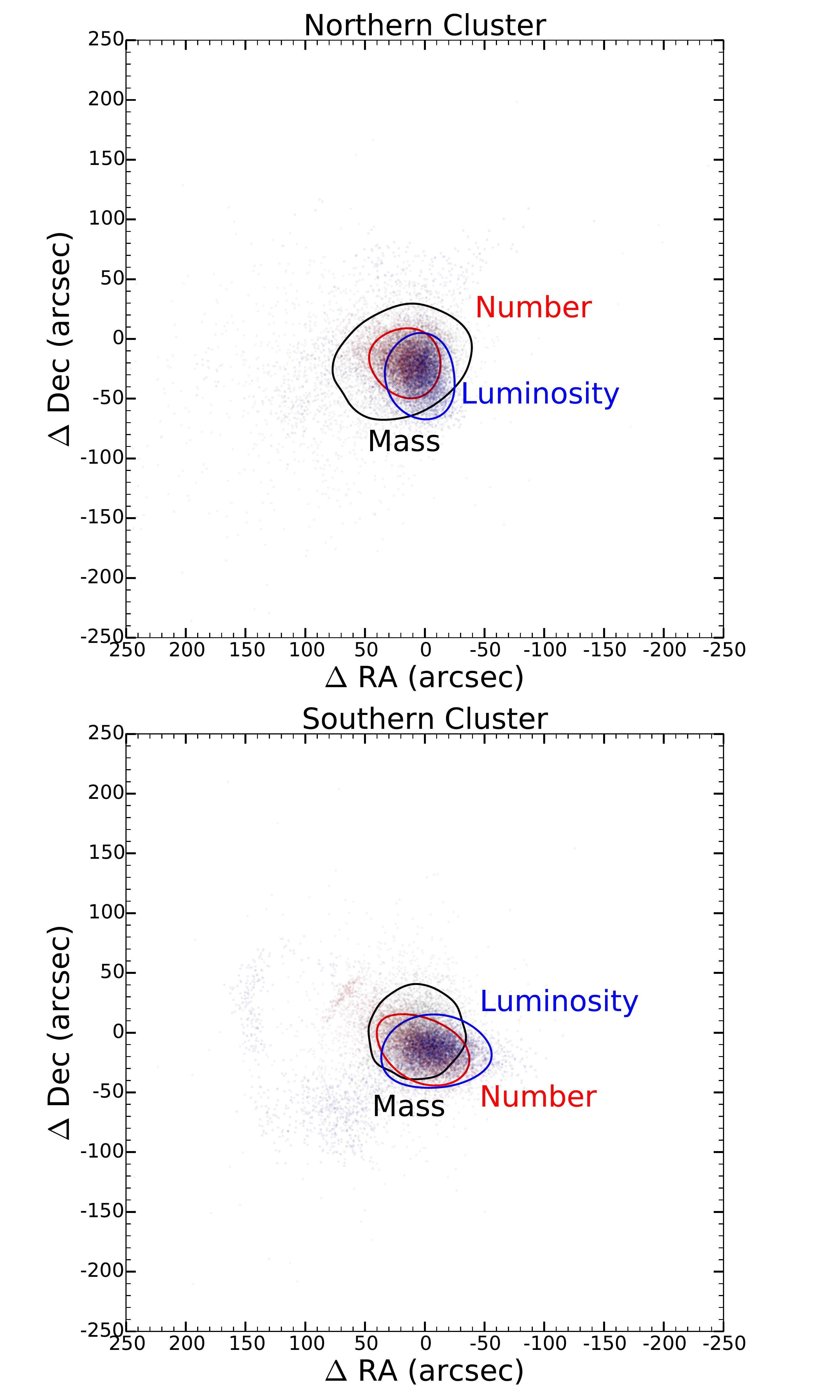}
        \caption{Centroid uncertainty estimation from 5000 bootstrap resampling runs. The coordinate (0,0) represents the peak of the mass centroid distribution. Black, red, and blue dots represent mass, number density, and luminosity peaks, respectively, from a single realization. The contours show the 1$\sigma$ confidence regions. } 
        \label{fig:peak_distribution}
\end{figure}

We used bootstrap analysis to measure the significance of the centroids for both mass and galaxy distributions. To estimate the WL mass distribution centroid uncertainty, we bootstrapped the final source catalog and generated 5000 convergence maps using the KS93 method. From each convergence map realization, we identified peaks by determining the first moment. Then, the distribution of the resulting peak locations was processed with a Kernel Density Estimation (KDE) to define the significance regions. We also generated 5000 bootstrap realizations of the galaxy distribution by sampling the photometrically and spectroscopically selected cluster members (presented in Figure \ref{fig:CMD}). Again, we identified peaks using the first moment and used KDE to define the significance regions.
While we believe that the resulting centroid uncertainty of the galaxy number density is a fair measure of the significance, we argue that the centroid uncertainty of the luminosity density obtained in this way corresponds to an upper limit because the peak location in each realization is dominated by several bright galaxies. Figure \ref{fig:peak_distribution} compares the 1$\sigma$ contours among the mass, luminosity, and number density results. We find that the three centroids are highly consistent with one another (well within 1$\sigma$ contours).

\subsection{Merging Scenario} 
\label{sec_merging_scenario}
A115 is a merging galaxy cluster with a number of intriguing features summarized as follows.
\begin{enumerate}
    \item A giant ($\mytilde2.5$~Mpc) radio relic is detected at the northern edge of A115N \citep{Govoni2001}.
    \item The orientation of the radio relic is approximately perpendicular to the vector connecting A115N and A115S.
    \item The center of the radio relic is offset toward the east by $\mytilde0.7$~Mpc from this connecting vector and $\mytilde1$~Mpc from the global center.
    \item Both surface brightness and temperature jumps in X-ray are detected across the radio relic, which is translated to $\mathcal{M}=1.4-2.0$ with systematics included \citep{Botteon2016}.
    \item The surface brightness distributions of the X-ray emission for both A115N and A115S are asymmetric. 
    \item Our WL analysis shows that A115S is twice more massive than  A115N and the total cluster mass is $M_{200c}=6.41_{-1.04}^{+1.08}\times10^{14}M_{\sun}$.
    \item The analysis with our enhanced spectroscopic catalog with 266 members shows that the LOS velocity difference between A115N and A115S is small ($244\pm144~\mbox{km}~\mbox{s}^{-1}$).
\end{enumerate}

Point 1 is strong evidence that the system is post merger. Although \cite{Hallman2018} suggests a possibility that the radio relic might be a pre-merger shock, our numerical simulation with our WL masses as input shows that this shock would be too weak to generate such a giant radio relic even if there exists a rich population of so-called fossil electrons (Lee et al. in prep). The last point supports the possibility that the merger is taking place nearly in the plane of the sky. Future radio observations can provide further insights into this viewing angle issue from polarization fraction measurements. Points 2 and 3 indicate that A115N and A115S might have collided in the north-south direction with a non-negligible impact parameter. Point 4 can be interpreted as suggesting that the shock velocity is as high as $\mytilde1800~\mbox{km}~\mbox{s}^{-1}$. Finally, we can infer from the morphology (Point 5) of the X-ray peak that A115N (A115S) might be heading southwest (northeast). 

Based on the subset of the points above, we can carry out some consistency checks for the progression of the merger. If the impact happened near the global center, the shock traveled $\mytilde1$~Mpc. Assuming a constant shock velocity of $\mytilde1800~\mbox{km}~\mbox{s}^{-1}$ derived from the Mach number, we estimate that it takes about 0.5 Gyr for the shock to reach the current location. Some simulations suggest that a shock traveling speed is a good proxy for the collision speed at the time of impact \citep[e.g.,][]{Springel2007}. 
An independent estimation of the collision velocity can be made with the following equation based on the so-called timing argument \citep{Sarazin2002}:
\begin{equation}\label{eq:merger_velocity}
\begin{aligned}
    v \mytilde 2930 &\left( \frac{M_1 + M_2}{10^{15}\ \text{M}_\odot} \right)^{1/2} \times \\ &\left( \frac{1 - d/d_0}{1-(b/d_0)^2} \right)^{1/2} \left(\frac{d}{1\ \text{Mpc}} \right)^{-1/2} \text{km s}^{-1},
\end{aligned}
\end{equation}
where the initial separation is
\begin{equation}
    d_0\  \mytilde \ 4.5 \left(\frac{M_1 + M_2}{10^{15}\ \text{M}_\odot} \right)^{1/3} \left(\frac{t_{\text{impact}}} {10\ \text{Gyr}} \right)^{2/3} \text{Mpc}.
\end{equation}
\noindent
We assume that $M_1+M_2=4.73\times10^{14}\ \text{M}_\odot$ is the sum of the individual WL cluster masses, $t_{\text{impact}}=11\ \text{Gyr}$ is the time from rest to impact, $b=0$ is the impact parameter, and the current separation $d=1$ Mpc. Setting $b=0$ is justified in this approximation because Equation \ref{eq:merger_velocity} varies quite slowly in $b/d_0$ when $d_0$ is large. The resulting relative velocity of the clusters at impact is $\mytilde 1700$ km s$^{-1}$. This agrees with the velocity derived from the Mach number of the shock. Furthermore, since the radio relic is close to A115N, it is unlikely that the subclusters have turned around and we are witnessing a returning phase. This is in contrast to the scenario that one might derive from the X-ray morphology.

More specific merger scenarios can be inferred when
we search for merging cluster analogs in cosmological numerical simulations. Using the  \cite{Wittman2018} method, 
we sampled cluster mergers by matching the cluster redshift, projected distance, radial velocity difference, and cluster masses.
As explained in \cite{Wittman2018}, this method has several 
advantages over the Monte Carlo Merger Analysis Code \citep[MCMAC;][]{dawson2013} method. One important advantage is that 
finding merger analogs in cosmological simulations
allows us to consider the cases where the subcluster velocity vectors are not entirely parallel to the separation vector 
while the MCMAC method always assumes that the collision is head-on.
This head-on collision assumption leads to on average a larger deviation between the separation vectors and the plane of the sky. As A115 is believed to be an off-axis merger, this issue cannot be neglected.

\begin{figure}
        \centering
        \includegraphics[width=0.46\textwidth]{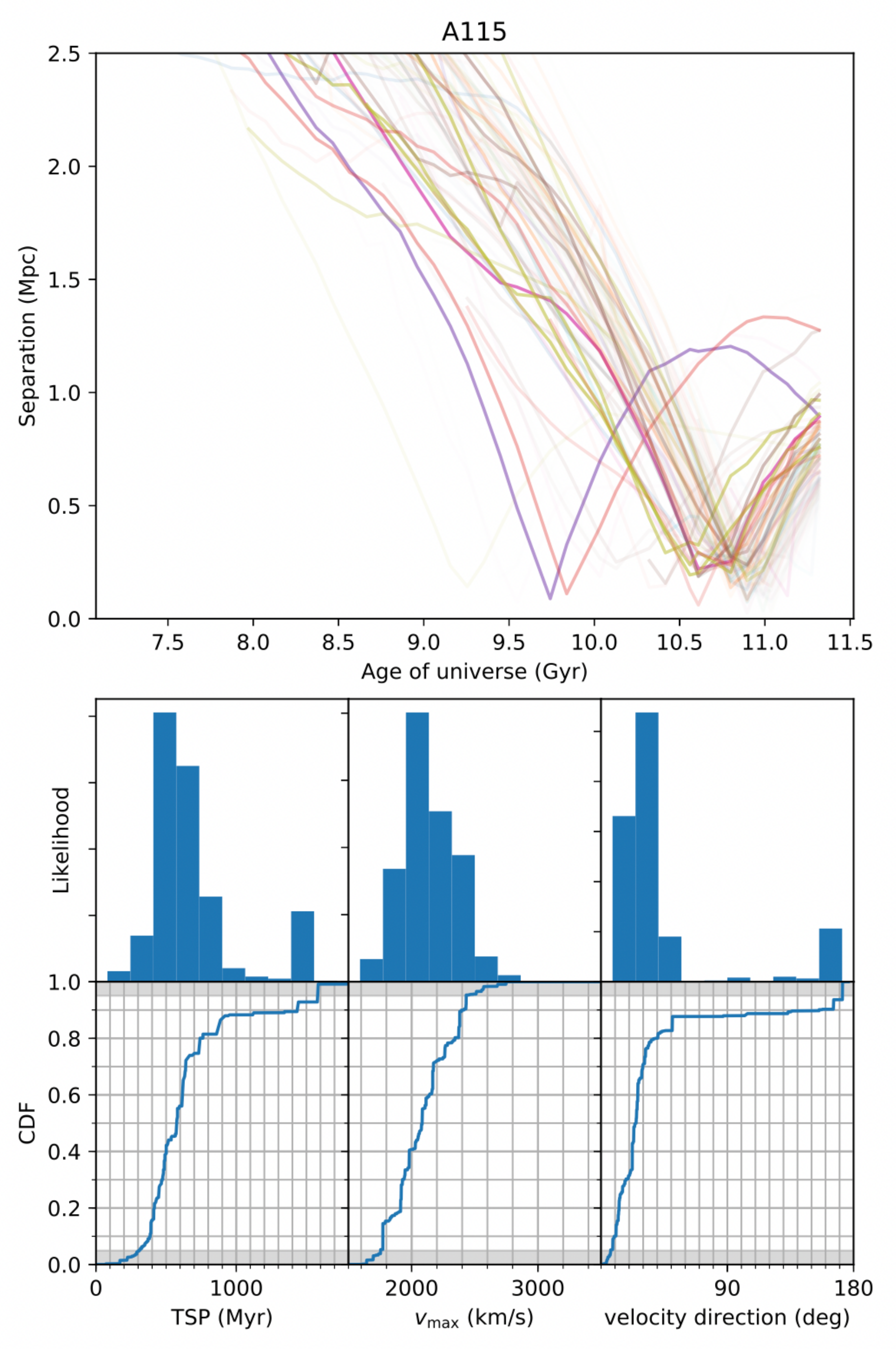}
        \caption{Trajectories of each cluster analog (top) and constraints of time since pericenter (TSP), maximum colliding velocity, and relative velocity direction with respect to the separation vector (bottom). Bottom panels present the likelihood and cumulative likelihood distributions. TSP is most likely to be \mytilde600 Myrs with a maximum colliding velocity of \mytilde2000 $\mbox{km}~\mbox{s}^{-1}$ and the relative velocity direction of \mytilde$25^{\circ}$ from the separation vector. This analysis prefers outgoing phase of subclusters, which is contradictory to what we would expect intuitively from the X-ray morphology. Note that the small bump near 160$\degr$ in the velocity direction panel (or near $\sim1500$ Myrs in the TSP panel) shows that a small fraction of the analogs are in the returning phase.} 
        \label{fig:merging_info}
\end{figure}

Figure~\ref{fig:merging_info} shows the trajectories of each analog (top) and constraints of time since pericenter (TSP; bottom left), maximum colliding velocity (bottom middle), and velocity direction (bottom right). 
TSP is a useful quantity because the information helps us to distinguish between in-bound and out-bound cases. The maximum colliding velocity is the impact velocity, which can be approximated to be the shock propagation velocity. Investigation of the velocity direction allows us to infer the viewing angle of the merger.
The time since pericenter is most likely to be $\mytilde600$ Myr with a maximum collision velocity of $\mytilde2000$~$\mbox{km}~\mbox{s}^{-1}$. These values are consistent with the above estimates based on the Mach number, position of the radio relic, and timing argument. The velocity direction (the angle between the relative velocity vector and the separation vector) is centered at $\mytilde25^{\circ}$. Although not shown here, we also found that about 68\% of analogs have their separation vector axis less than $19^{\circ}$ from the plane of the sky. Therefore, our LOS velocity difference constraint $244\pm144~\mbox{km}~\mbox{s}^{-1}$ only marginally favors mergers near the plane of the sky. This weak constraint is not surprising because the velocity vectors of the analogs are not perfectly aligned with the separation vectors.
The trajectory plot (top) shows that the majority of the analogs are in the outgoing phase at the cluster redshift. This can also be inferred by either the short TSP or the relative velocity vector being less than 90$\degr$; the relative velocity vector is (mostly) parallel to the separation vector, rather than anti-parallel. 
Since we do not use the radio relic in our analog search, it is interesting that this analog-based result also favors the same outgoing case. However, note that the small bump near 160$\degr$ in the velocity direction panel (or near $\sim1.5$~Gyrs in the TSP panel) shows that a small fraction of the analogs are in the returning phase.

In summary, we find that our analysis favors A115N and A115S in an outgoing phase. However, this seems to contradict the visual impression given by the cometary tails in X-ray emission. Careful hydrodynamical simulations are needed to determine whether or not the observed X-ray morphologies can be reproduced in an outgoing phase.

\section{Conclusions} \label{sec_conclusion}

A115 is a merging galaxy cluster with a number of remarkable features including a giant ($\mytilde2.5$ Mpc) radio relic and two asymmetric X-ray peaks with trailing tails. 
Having presented a detailed multi-wavelength analysis of A115 including imaging  data  from  Subaru, X-ray data from {\it Chandra}, spectroscopic data from the Keck/DEIMOS and MMT/Hectospec instruments, we summarize our conclusions as follows:
\begin{itemize}
    \item Our WL study confirms the finding of \cite{Okabe2010} that the mass structure of A115 is bimodal and resembles
    the X-ray map.
    \item Both mass clumps are in good spatial agreement with the distributions of galaxies and plasma.
    \item We determine the masses of A115N and A115S to be $M_{200c}=1.58^{+0.56}_{-0.49} \times10^{14}M_{\sun}$ and $M_{200c}=3.15_{-0.71}^{+0.79} \times10^{14}M_{\sun}$, respectively. The total mass of the system is $M_{200c}=6.41_{-1.04}^{+1.08} \times10^{14}M_{\sun}$.
    \item The mass estimates made with our X-ray and spectroscopic data analysis are 3-10 times higher than the WL values. We attribute the difference to severe disruption of the gravitational and hydrostatic structure due to the merger. When we adopt non-WL masses, the $M/L$ values of A115 become unphysically high.
    \item Our dynamical analysis of A115 with 266 cluster members shows that the LOS speed is low, suggesting a higher chance of the merger taking place nearly in the plane of sky. Our cluster analogs support this theory and constrain the separation vector to be greater than
    $\mytilde71\degr$ from the LOS (i.e., less than $\mytilde19\degr$ from the plane of the sky) 68\% of the time. Although we agree with \cite{Barrena2007} that the central galaxies around the BCG of A115N (including the BCG) tend to possess larger velocities than the mean value, the significance is low.
\end{itemize}

From our multi-wavelength data analysis, we suggest a scenario wherein we may be witnessing the outgoing phase of the cluster merger after first passage. However, detailed high-fidelity numerical simulations are required to draw a firm conclusion on the merger phase of A115. In particular, it will be interesting to investigate whether or not merger-induced dynamical disruptions can inflate the measured velocity dispersion and cause such a large discrepancy as our observations show.

\acknowledgments
{We thank Andrea Botteon for sharing his VLA radio data and Bill Forman, Wonki Lee, Cristiano Sabiu, and Ho Seong Hwang for useful discussions.
MJJ acknowledges support for the current research from the National Research Foundation of Korea under the program 2017R1A2B2004644 and 2017R1A4A1015178.
Portions of this work was performed under the auspices of the U.S. Department of Energy by Lawrence Livermore National Laboratory under contract DE-AC52-07NA27344. Lawrence Livermore National Security, LLC. RJvW acknowledges support from the VIDI research programme with project number 639.042.729, which is financed by the Netherlands Organisation for Scientific Research (NWO).
}

\bibliographystyle{aasjournal.bst}

\begin{thebibliography}{}

\bibitem[Ahn et al.(2012)]{Ahn2012} Ahn, C.~P., Alexandroff, R., Allende Prieto, C., et al.\ 2012, \apjs, 203, 21

\bibitem[Alam et al.(2015)]{Alam2015} Alam, S., Albareti, F.~D., Allende Prieto, C., et al.\ 2015, \apjs, 219, 12

\bibitem[Barrena et al.(2007)]{Barrena2007} Barrena, R., Boschin, W., Girardi, M., \& Spolaor, M.\ 2007, \aap, 469, 861

\bibitem[Beers et al.(1983)]{Beers1983} Beers, T.~C., Huchra, J.~P., \& Geller, M.~J.\ 1983, \apj, 264, 356 

\bibitem[Beers et al.(1990)]{Beers1990} Beers, T.~C., Flynn, K., \& Gebhardt, K.\ 1990, \aj, 100, 32 

\bibitem[Ben{\'{\i}}tez et al.(2009)]{Benitez2009} Ben{\'{\i}}tez, N., Moles, M., Aguerri, J.~A.~L., et al.\ 2009, \apjl, 692, L5

\bibitem[Bertin \& Arnouts(1996)]{Bertin1996} Bertin, E., \& Arnouts, S.\ 1996, \aaps, 117, 393 

\bibitem[Bertin et al.(2002)]{Bertin2002} Bertin, E., Mellier, Y., Radovich, M., et al.\ 2002, Astronomical Data Analysis Software and Systems XI, 281, 228

\bibitem[Bertin(2006)]{Bertin2006} Bertin, E.\ 2006, Astronomical Data Analysis Software and Systems XV, 351, 112

\bibitem[Bonafede et al.(2014)]{bonafede2014} Bonafede, A., Intema, H.~T., Br{\"u}ggen, M., et al.\ 2014, \apj, 785, 1 

\bibitem[Botteon et al.(2016)]{Botteon2016} Botteon, A., Gastaldello, F., Brunetti, G., \& Dallacasa, D.\ 2016, \mnras, 460, L84

\bibitem[Br{\"u}ggen et al.(2011)]{bruggen2011} Br{\"u}ggen, M., van Weeren, R.~J., \& R{\"o}ttgering, H.~J.~A.\ 2011, \memsai, 82, 627 

\bibitem[Carlberg et al.(1997)]{carlberg1997} Carlberg, R.~G., Yee, H.~K.~C., Ellingson, E., et al.\ 1997, \apjl, 485, L13 

\bibitem[Chartas \& Getman(2002)]{chartas2002}Chartas, G., \& Getman, K. 2002, ACISABS: The ACIS Time Dependent Absorption due to Molecular Contamination, https://heasarc.gsfc.nasa.gov/xanadu/xspec/models/acisabs.html

\bibitem[Clowe et al.(1998)]{Clowe1998} Clowe, D., Luppino, G.~A., Kaiser, N., Henry, J.~P., \& Gioia, I.~M.\ 1998, \apjl, 497, L61

\bibitem[Condon et al.(1998)]{Condon1998} Condon, J.~J., Cotton, W.~D., Greisen, E.~W., et al.\ 1998, \aj, 115, 1693

\bibitem[Dahlen et al.(2010)]{Dahlen2010} Dahlen, T., Mobasher, B., Dickinson, M., et al.\ 2010, \apj, 724, 425 

\bibitem[Dawson(2013)]{dawson2013} Dawson, W.~A.\ 2013, \apj, 772, 131 

\bibitem[Dempster et al.(1977)]{Dempster1977} Dempster, Arthur P and Laird, Nan M and Rubin, Donald B, et al.\ 1977, Journal of the royal statistical society. Series B (methodological), 1--38 

\bibitem[Di Gennaro et al.(2018)]{Gennaro2018} Di Gennaro, G., van Weeren, R.~J., Hoeft, M., et al.\ 2018, \apj, 865, 24

\bibitem[Dutton \& Macci{\`o}(2014)]{Dutton2014} Dutton, A.~A., \& Macci{\`o}, A.~V.\ 2014, \mnras, 441, 3359

\bibitem[Fahlman et al.(1994)]{Fahlman1994} Fahlman, G., Kaiser, N., Squires, G., \& Woods, D.\ 1994, \apj, 437, 56

\bibitem[Ferrari et al.(2008)]{Ferrari2008} Ferrari, C., Govoni, F., Schindler, S., Bykov, A.~M., \& Rephaeli, Y.\ 2008, \ssr, 134, 93 

\bibitem[Finner et al.(2017)]{Finner2017} Finner, K., Jee, M.~J., Golovich, N., et al.\ 2017, \apj, 851, 46

%\bibitem[Finoguenov et al.(2001)]{Finoguenov2001} Finoguenov, A., Reiprich, T.~H., \& B{\"o}hringer, H.\ 2001, \aap, 368, 749

\bibitem[Forman et al.(1981)]{Forman1981} Forman, W., Bechtold, J., Blair, W., et al.\ 1981, \apjl, 243, L133

%\bibitem[Girardi et al.(2000)]{Girardi2000} Girardi, M., Borgani, S., Giuricin, G., Mardirossian, F., \& Mezzetti, M.\ 2000, \apj, 530, 62

\bibitem[Giavalisco et al.(2004)]{Giavalisco2004} Giavalisco, M., Ferguson, H.~C., Koekemoer, A.~M., et al.\ 2004, \apjl, 600, L93 

\bibitem[Girardi et al.(2002)]{Girardi2002} Girardi, M., Manzato, P., Mezzetti, M., Giuricin, G., \& Limboz, F.\ 2002, \apj, 569, 720 

\bibitem[Golovich et al.(2017)]{Golovich2017} Golovich, N., Dawson, W.~A., Wittman, D.~M., et al.\ 2017, arXiv:1711.01347

\bibitem[Golovich et al.(2018)]{Golovich2018} Golovich, N., Dawson, W.~A., Wittman, D.~M., et al.\ 2018, arXiv:1806.10619 

\bibitem[Govoni et al.(2001)]{Govoni2001} Govoni, F., Feretti, L., Giovannini, G., et al.\ 2001, \aap, 376, 803

\bibitem[Guo et al.(2013)]{Guo2013} Guo, Y., Ferguson, H.~C., Giavalisco, M., et al.\ 2013, \apjs, 207, 24

\bibitem[Gutierrez \& Krawczynski(2005)]{Gutierrez2005} Gutierrez, K., \& Krawczynski, H.\ 2005, \apj, 619, 161

\bibitem[Hallman et al.(2018)]{Hallman2018} Hallman, E.~J., Alden, B., Rapetti, D., Datta, A., \& Burns, J.~O.\ 2018, \apj, 859, 44

\bibitem[Hoang et al.(2018)]{Hoang2018} Hoang, D.~N., Shimwell, T.~W., van Weeren, R.~J., et al.\ 2018, arXiv:1811.09713

\bibitem[Hoekstra et al.(2000)]{Hoekstra2000} Hoekstra, H., Franx, M., \& Kuijken, K.\ 2000, \apj, 532, 88 

\bibitem[Hoekstra et al.(2012)]{Hoekstra2012} Hoekstra, H., Mahdavi, A., Babul, A., \& Bildfell, C.\ 2012, \mnras, 427, 1298 

\bibitem[Jee et al.(2005)]{Jee2005} Jee, M.~J., White, R.~L., Ben{\'{\i}}tez, N., et al.\ 2005, \apj, 618, 46

\bibitem[Jee et al.(2007)]{Jee2007} Jee, M.~J., Blakeslee, J.~P., Sirianni, M., et al.\ 2007, \pasp, 119, 1403 

\bibitem[Jee et al.(2007)]{Jee2007b} Jee, M.~J., Ford, H.~C., Illingworth, G.~D., et al.\ 2007, \apj, 661, 728 

\bibitem[Jee \& Tyson(2011)]{JeeTyson2011} Jee, M.~J., \& Tyson, J.~A.\ 2011, \pasp, 123, 596 

\bibitem[Jee et al.(2013)]{Jee2013} Jee, M.~J., Tyson, J.~A., Schneider, M.~D., et al.\ 2013, \apj, 765, 74

\bibitem[Jee et al.(2014)]{Jee2014} Jee, M.~J., Hoekstra, H., Mahdavi, A., \& Babul, A.\ 2014, \apj, 783, 78

\bibitem[Jee et al.(2015)]{Jee2015} Jee, M.~J., Stroe, A., Dawson, W., et al.\ 2015, \apj, 802, 46

\bibitem[Jee et al.(2016)]{Jee2016} Jee, M.~J., Dawson, W.~A., Stroe, A., et al.\ 2016, \apj, 817, 179 

\bibitem[Jester et al.(2005)]{Jester2005} Jester, S., Schneider, D.~P., Richards, G.~T., et al.\ 2005, \aj, 130, 873

\bibitem[Kaastra \& Mewe(1993)]{Kaastra1993} Kaastra, J.~S., \& Mewe, R.\ 1993, \aaps, 97, 443

\bibitem[Kaiser \& Squires(1993)]{Kaiser1993} Kaiser, N., \& Squires, G.\ 1993, \apj, 404, 441

\bibitem[Kaiser et al.(1995)]{Kaiser1995} Kaiser, N., Squires, G., \& Broadhurst, T.\ 1995, \apj, 449, 460

\bibitem[Kang \& Ryu(2011)]{Kang2011} Kang, H., \& Ryu, D.\ 2011, \apj, 734, 18 

\bibitem[Kang et al.(2012)]{Kang2012} Kang, H., Ryu, D., \& Jones, T.~W.\ 2012, \apj, 756, 97 

\bibitem[Kang \& Ryu(2015)]{Kang2015} Kang, H., \& Ryu, D.\ 2015, \apj, 809, 186 

\bibitem[Landry et al.(2013)]{Landry2013} Landry, D., Bonamente, M., Giles, P., et al.\ 2013, \mnras, 433, 2790 

\bibitem[Lidman et al.(2012)]{Lidman2012} Lidman, C., Suherli, J., Muzzin, A., et al.\ 2012, \mnras, 427, 550

\bibitem[Liedahl et al.(1995)]{Liedahl1995} Liedahl, D.~A., Osterheld, A.~L., \& Goldstein, W.~H.\ 1995, \apjl, 438, L115

\bibitem[Mahdavi et al.(2013)]{Mahdavi2013} Mahdavi, A., Hoekstra, H., Babul, A., et al.\ 2013, \apj, 767, 116 

\bibitem[Mandelbaum et al.(2015)]{Mandelbaum2015} Mandelbaum, R., Rowe, B., Armstrong, R., et al.\ 2015, \mnras, 450, 2963 

\bibitem[Mantz et al.(2016)]{Mantz2016} Mantz, A.~B., Allen, S.~W., Morris, R.~G., et al.\ 2016, \mnras, 463, 3582

\bibitem[Markevitch et al.(2004)]{Markevitch2004} Markevitch, M., Gonzalez, A.~H., Clowe, D., et al.\ 2004, \apj, 606, 819

\bibitem[Miyazaki et al.(2002)]{Miyazaki2002} Miyazaki, S., Komiyama, Y., Sekiguchi, M., et al.\ 2002, \pasj, 54, 833 

\bibitem[Monteiro-Oliveira et al.(2017)]{Monteiro-Oliveira2017} Monteiro-Oliveira, R., Cypriano, E.~S., Machado, R.~E.~G., et al.\ 2017, \mnras, 466, 2614 

\bibitem[Navarro et al.(1997)]{Navarro1997} Navarro, J.~F., Frenk, C.~S., \& White, S.~D.~M.\ 1997, \apj, 490, 493 

\bibitem[Niemi et al.(2015)]{Niemi2015} Niemi, S.-M., Kitching, T.~D., \& Cropper, M.\ 2015, \mnras, 454, 1221

\bibitem[Niikura et al.(2015)]{Niikura2015} Niikura, H., Takada, M., Okabe, N., Martino, R., \& Takahashi, R.\ 2015, \pasj, 67, 103 

\bibitem[Oguri et al.(2010)]{Oguri2010} Oguri, M., Takada, M., Okabe, N., \& Smith, G.~P.\ 2010, \mnras, 405, 2215

\bibitem[Okabe et al.(2010)]{Okabe2010} Okabe, N., Takada, M., Umetsu, K., Futamase, T., \& Smith, G.~P.\ 2010, \pasj, 62, 811

\bibitem[Ouchi et al.(2004)]{Ouchi2004} Ouchi, M., Shimasaku, K., Okamura, S., et al.\ 2004, \apj, 611, 660

\bibitem[Pedersen \& Dahle(2007)]{Pedersen2007} Pedersen, K., \& Dahle, H.\ 2007, \apj, 667, 26

\bibitem[Pinkney et al.(1996)]{Pinkney1996} Pinkney, J., Roettiger, K., Burns, J.~O., \& Bird, C.~M.\ 1996, \apjs, 104, 1

\bibitem[Pinzke et al.(2013)]{Pinzke2013} Pinzke, A., Oh, S.~P., \& Pfrommer, C.\ 2013, \mnras, 435, 1061 

\bibitem[Ragozzine et al.(2012)]{Ragozzine2012} Ragozzine, B., Clowe, D., Markevitch, M., Gonzalez, A.~H., \& Brada{\v c}, M.\ 2012, \apj, 744, 94

\bibitem[Randall et al.(2008)]{Randall2008} Randall, S.~W., Markevitch, M., Clowe, D., Gonzalez, A.~H., \& Brada{\v c}, M.\ 2008, \apj, 679, 1173 

\bibitem[Rines et al.(2018)]{Rines2018} Rines, K.~J., Geller, M.~J., Diaferio, A., Hwang, H.~S., \& Sohn, J.\ 2018, \apj, 862, 172

\bibitem[Sarazin(2002)]{Sarazin2002} Sarazin, C.~L.\ 2002, Merging Processes in Galaxy Clusters, 272, 1 

\bibitem[Saro et al.(2013)]{Saro2013} Saro, A., Mohr, J.~J., Bazin, G., \& Dolag, K.\ 2013, \apj, 772, 47

\bibitem[Schlegel et al.(1998)]{Schlegel1998} Schlegel, D.~J., Finkbeiner, D.~P., \& Davis, M.\ 1998, \apj, 500, 525

\bibitem[Seitz \& Schneider(1997)]{Seitz1997} Seitz, C., \& Schneider, P.\ 1997, \aap, 318, 687

\bibitem[Shibata et al.(1999)]{Shibata1999} Shibata, R., Honda, H., Ishida, M., Ohashi, T., \& Yamashita, K.\ 1999, \apj, 524, 603

\bibitem[Sif{\'o}n et al.(2015)]{Sifon2015} Sif{\'o}n, C., Hoekstra, H., Cacciato, M., et al.\ 2015, \aap, 575, A48

\bibitem[Sirianni et al.(2005)]{Sirianni2005} Sirianni, M., Jee, M.~J., Ben{\'{\i}}tez, N., et al.\ 2005, \pasp, 117, 1049

\bibitem[Skillman et al.(2013)]{Skillman2013} Skillman, S.~W., Xu, H., Hallman, E.~J., et al.\ 2013, \apj, 765, 21 

\bibitem[Skrutskie et al.(2006)]{Skrutskie2006} Skrutskie, M.~F., Cutri, R.~M., Stiening, R., et al.\ 2006, \aj, 131, 1163

\bibitem[Soucail(2012)]{Soucail2012} Soucail, G.\ 2012, \aap, 540, A61 

\bibitem[Springel \& Farrar(2007)]{Springel2007} Springel, V., \& Farrar, G.~R.\ 2007, \mnras, 380, 911 

\bibitem[Stark et al.(1992)]{Stark1992} Stark, A.~A., Gammie, C.~F., Wilson, R.~W., et al.\ 1992, \apjs, 79, 77

\bibitem[Stroe et al.(2014)]{Stroe2014} Stroe, A., Harwood, J.~J., Hardcastle, M.~J., \& R{\"o}ttgering, H.~J.~A.\ 2014, \mnras, 445, 1213

\bibitem[Takizawa et al.(2010)]{Takizawa2010} Takizawa, M., Nagino, R., \& Matsushita, K.\ 2010, \pasj, 62, 951

\bibitem[Urdampilleta et al.(2018)]{Urdampilleta2018} Urdampilleta, I., Akamatsu, H., Mernier, F., et al.\ 2018, \aap, 618, A74

\bibitem[van Weeren et al.(2017)]{vanWeeren2017} van Weeren, R.~J., Andrade-Santos, F., Dawson, W.~A., et al.\ 2017, Nature Astronomy, 1, 0005

\bibitem[Vazza et al.(2012)]{Vazza2012} Vazza, F., Br{\"u}ggen, M., van Weeren, R., et al.\ 2012, \mnras, 421, 1868

\bibitem[White et al.(1997)]{White1997} White, D.~A., Jones, C., \& Forman, W.\ 1997, \mnras, 292, 419

\bibitem[Wittman et al.(2018)]{Wittman2018} Wittman, D., Cornell, B.~H., \& Nguyen, J.\ 2018, \apj, 862, 160 

\bibitem[Wright \& Brainerd(2000)]{Wright2000} Wright, C.~O., \& Brainerd, T.~G.\ 2000, \apj, 534, 34 

\bibitem[Yagi et al.(2002)]{Yagi2002} Yagi, M., Kashikawa, N., Sekiguchi, M., et al.\ 2002, \aj, 123, 66

\bibitem[Zabludoff et al.(1990)]{Zabludoff1990} Zabludoff, A.~I., Huchra, J.~P., \& Geller, M.~J.\ 1990, \apjs, 74, 1

\bibitem[Zhang et al.(2010)]{Zhang2010} Zhang, Y.-Y., Okabe, N., Finoguenov, A., et al.\ 2010, \apj, 711, 1033 

\end{thebibliography}

\end{document}